%% file: main.tex
\PassOptionsToPackage{pdfa}{hyperref}
\documentclass[sigconf, nonacm]{acmart}

\usepackage{pvldb}

\renewcommand\vldbdoi{10.14778/3860016.3860018}
\renewcommand\vldbpages{481 - 495}
\renewcommand\vldbissue{3}
\renewcommand\vldbyear{2026}
\renewcommand\vldbavailabilityurl{https://github.com/AIS-SNU/NAVIS}
\renewcommand\vldbpagestyle{empty}

\input{others/packages}
\input{others/commands}

\begin{document}

\title{\thiswork: Concurrent Search and Update with Low Position-Seeking Overhead in On-SSD Graph-Based Vector Search}

\input{others/authors}

\input{sections/abstract}

\maketitle

\vldbtopmatter

\input{sections/intro}
\input{sections/background}
\input{sections/motivation}

\input{sections/overview}
\input{sections/method_layout}
\input{sections/method_entrance}
\input{sections/method_cache}
\input{sections/impl}

\input{sections/eval}
\input{sections/discussion}
\input{sections/related_work}
\input{sections/conclusion}

\begin{acks}
We thank the reviewers for their invaluable feedback.
Jinho Lee is the corresponding author.
This work was primarily supported by 
Samsung Electronics. 
Jinho Lee is additionally funded by 
National Research Foundation of Korea (NRF) grant funded by the Korea government (MSIT) (RS-2026-25495605), 
and 
Institute of Information \& communications Technology Planning \& Evaluation (IITP) 
(RS-2024-00395134, 
RS-2024-00347394, 
RS-2026-25548502, 
RS-2026-25549926, 
RS-2021II211343).   
Part of the infrastructure used in this work was supported by Korea Basic Science Institute (National Research Facilities and Equipment Center) grant funded by the Ministry of Science and ICT (No. RS-2025-00564840). 
Jaeyong Song was additionally supported by Information \& communications Technology Planning \& Evaluation (IITP) under the Graduate School of Artificial Intelligence Semiconductor (IITP-2026-RS-2023-00256081) grant funded by the Korea government (MSIT).
\end{acks}

\bibliographystyle{ACM-Reference-Format}
\bibliography{refs}

\end{document}

%% file: others/packages.tex
\usepackage{tikz}
\newcommand*\circled[1]{\tikz[baseline=(char.base)]{
            \node[shape=circle,draw,inner sep=0.4pt, fill=white, text=black] (char) {#1};}}

\newcommand*\bcircled[1]{\tikz[baseline=(char.base)]{
            \node[shape=circle,inner sep=0.4pt, fill=black, text=white] (char) {#1};}}

\usepackage{algorithm}
\usepackage{amsmath}
\usepackage{algpseudocode}

\usepackage[capitalize]{cleveref}

\usepackage{xspace}
\usepackage{enumitem}

\usepackage{makecell}
\usepackage{booktabs}
\usepackage{multirow}

\hypersetup{pdfapart=2, pdfaconformance=B}

%% file: others/commands.tex
\newcommand{\thiswork}{\textsc{NAVIS}\xspace}

\newcommand{\layout}{locality-driven decoupling\xspace}
\newcommand{\Layout}{Locality-driven decoupling\xspace}
\newcommand{\LayOut}{Locality-Driven Decoupling\xspace}

\newcommand{\vecpruning}{convergence-aware speculative reranking\xspace}
\newcommand{\Vecpruning}{Convergence-aware speculative reranking\xspace}
\newcommand{\VecPruning}{Convergence-Aware Speculative Reranking\xspace}

\newcommand{\navgraph}{entrance graph\xspace}
\newcommand{\NavGraph}{Entrance Graph\xspace}

\newcommand{\cache}{\thiswork-cache\xspace}
\newcommand{\Cache}{\thiswork-cache\xspace}
\newcommand{\CaChe}{\thiswork-Cache\xspace}

\newcommand{\layoutpruning}{\thiswork-reader\xspace}
\newcommand{\LayoutPruning}{\thiswork-Reader\xspace}

\newcommand{\navupdate}{\thiswork-update\xspace}
\newcommand{\Navupdate}{\thiswork-update\xspace}
\newcommand{\NavUpdate}{\thiswork-Update\xspace}

\newcommand{\Naive}{Na\"ive\xspace}
\newcommand{\naive}{na\"ive\xspace}

\newcommand{\FDANN}{\textsc{FreshDiskANN}\xspace}
\newcommand{\IPDANN}{\textsc{IP-DiskANN}\xspace}
\newcommand{\ODANN}{\textsc{OdinANN}\xspace}
\newcommand{\ODANNP}{\textsc{OdinANN+\$}\xspace}
\newcommand{\Qdrant}{\textsc{Qdrant}\xspace}
\newcommand{\SPFresh}{\textsc{SPFresh}\xspace}

\newcommand{\fineweb}{\texttt{FineWeb}\xspace}
\newcommand{\msmarco}{\texttt{MSMARCO}\xspace}
\newcommand{\deep}{\texttt{DEEP}\xspace}
\newcommand{\imgnet}{\texttt{ImageNet}\xspace}

%% file: others/authors.tex
\author{Jaeyong Song}
\authornote{Work partially done while at University of Southern California.}
\affiliation{%
  \institution{Seoul National University}
}
\email{jaeyong.song@snu.ac.kr}

\author{Hongsun Jang}
\affiliation{%
  \institution{Seoul National University}
}
\email{hongsun.jang@snu.ac.kr}

\author{Changmin Shin}
\affiliation{%
  \institution{Seoul National University}
}
\email{scm8432@snu.ac.kr}

\author{Seongyeon Park}
\affiliation{%
  \institution{Seoul National University}
}
\email{syeonp@snu.ac.kr}

\author{Yong Jae Ryoo}
\affiliation{%
  \institution{Seoul National University}
}
\email{jae8259@snu.ac.kr}

\author{Seo Jin Park}
\affiliation{%
  \institution{University of Southern California}
}
\email{seojin.park@usc.edu}

\author{Jinho Lee}
\affiliation{%
  \institution{Seoul National University}
}
\email{leejinho@snu.ac.kr}

\settopmatter{authorsperrow=4}

\makeatletter
\def\@ACM@checkaffil{}
\newcount\vldbauthorboxcount
\let\vldbOriginalAuthorBox\@typeset@author@bx

\def\@typeset@author@bx{%
  \advance\vldbauthorboxcount by 1\relax
  \ifnum\vldbauthorboxcount=5\relax
    \author@bx@wd=\dimexpr
      (\textwidth-\author@bx@sep)/3
      -\author@bx@sep\relax
  \fi
  \vldbOriginalAuthorBox
}
\makeatother

%% file: sections/abstract.tex
\begin{abstract}
On-disk graph-based vector search (GVS) has become the dominant approach for serving large-scale vector databases at high recall.
Yet prior systems struggle to sustain concurrent search and update throughput on high-dimensional workloads.
We identify the main bottleneck as \textit{position seeking}, the graph traversal that finds neighbors during each update.
Unlike search, which returns only the top-$K$ closest vectors, it explores a much larger set to build an $R$-neighbor list of close neighbors and long-range shortcuts.
These neighbors have different distance-accuracy requirements, yet search-oriented packed layouts fetch full vectors with every edgelist access, consuming cache capacity regardless of their role.
We present \thiswork, an on-SSD GVS system that redesigns position seeking around this mismatch.
At its core, \layout separates vectors from edgelists and packs graph-local edgelists together.
This decoupling enables \vecpruning to selectively load full vectors until a sufficient set of close neighbors is resolved, while selecting the remaining shortcuts using approximate distances.
It also allows \thiswork to devote cache capacity entirely to edgelists.
To make this cache effective, \thiswork keeps the \navgraph fresh via lightweight updates that reuse traversal information already produced during position seeking, thereby maintaining a stable hot set of edgelists near its entry points.
Across multiple large-scale benchmarks, \thiswork improves insertion throughput by up to 3.97$\times$ and concurrent search throughput by up to 4.00$\times$ while reducing search latency by up to 75.69\%.
\end{abstract}

%% file: sections/intro.tex
\section{Introduction}
\label{sec:intro}

Modern applications like retrieval-augmented generation~\cite{rag}, recommendation systems~\cite{rec1, rec2}, and multimedia search~\cite{multimedia1, multimedia2} increasingly rely on large-scale vector databases to retrieve semantically similar items at high recall.
Among various designs, on-disk \textit{graph-based vector search} (GVS)~\cite{diskann,starling,pipeann,freshdiskann,odinann,qdrant} has emerged as the dominant approach for high-recall search at scale.
In on-disk GVS, vectors are vertices in a proximity graph whose edges link nearby points, and a query greedily traverses from a starting vertex toward progressively closer neighbors, loading only the visited portions of the index from SSD.
Recent production deployments~\cite{qdrant, faiss} adopt on-disk GVS since it sustains high recall without holding full vectors in DRAM~\cite{youtube,jdcom}, especially for the 512 to 3072-dimensional vectors that modern encoders produce~\cite{clip, ada002, openai_embedding, bge_m3}.

In practice, vector databases are continuously updated after deployment.
With the rise of agentic systems~\cite{rag-update1, rag-update2}, and as production pipelines~\cite{recom1, recom2, youtube} now ingest fresh user activity and frequent uploads, on-disk GVS systems are no longer static snapshots.
Such updates are handled either by \emph{index rebuild} or by \emph{index update}.
As a rebuild often takes several days at large scale~\cite{diskann}, many approaches~\cite{freshdiskann, odinann, ipdiskann, greator} adopt index update to incorporate fresh vectors quickly.
However, sustaining high search performance under such concurrent index updates is non-trivial.
Despite several proposals for index updates~\cite{freshdiskann, greator, ipdiskann, odinann},
they often fail to sustain high concurrent search and update throughput that modern high-dimensional workloads demand.
These systems improve how updates are incorporated or committed, but largely retain the search-oriented traversal used to find neighbors before each update.

We find that \textbf{\emph{position seeking}} is the main bottleneck for index update.
Position seeking traverses the graph to find neighbors for the newly inserted vertices.
This process resembles a search query, but it must explore the graph more extensively to identify a large set of neighbors for graph construction.
Specifically, a search query stops once it collects the top-$K$ (e.g., 10) closest vectors, whereas an index update must gather enough neighbors to match the graph's out-degree $R$ (e.g., 64--96) to preserve graph quality.
Consequently, position seeking must read far more vectors, leading to costly SSD accesses.
Especially on high-dimensional vector databases, position seeking consumes up to 85\% of update time and cuts concurrent search throughput by 27.89\% on average (\cref{sec:moti}).

A closer look reveals that much of this overhead stems from design choices inherited from search-oriented frameworks.
Existing frameworks commonly adopt a \emph{packed layout}~\cite{starling, odinann, freshdiskann}, which stores each vector together with its edgelist on the same SSD page.
This layout suits search workloads, since each visited vertex typically requires both its vector for distance computation and its edgelist for further traversal.
For position seeking, however, we find that this design introduces two key inefficiencies.

First, search and position seeking impose different distance-accuracy requirements, but the packed layout serves both with the same full-vector I/O.
A search query requires accurately distinguishing its top-$K$ results.
Instead, position seeking constructs an $R$-neighbor list whose members serve complementary roles: close neighbors preserve local connectivity, while long-range shortcuts improve graph navigability and reduce traversal I/O~\cite{diskann}.
These long-range shortcuts do not require the same degree of distance discrimination as the close neighbors~\cite{freshdiskann}.
Nevertheless, accessing their edgelists under the packed layout brings in their full vectors as well, resulting in unnecessary vector reads.

Second, the packed layout exacerbates the cache pressure inherent in position seeking.
To identify $R$ neighbors, position seeking explores substantially more vertices than a search query that returns only the top-$K$, resulting in a much larger working set.
This pressure further increases as vector databases grow in both the number and dimensionality of vectors, making it increasingly difficult to retain the working set in memory~\cite{gorgeous}.
One possibility is to selectively cache edgelists, avoiding the management of high-dimensional vectors in cache.
However, the packed layout makes such a choice inefficient: caching entire pages wastes cache capacity on vectors, thereby limiting the effective cache size.

From the above observations,
we present \emph{\thiswork}, an on-SSD GVS system that redesigns position seeking around \emph{\layout}, which separates vectors from edgelists so a traversal loads edgelists without dragging in vectors and an update writes only the edgelists it changes.
Freed of vectors, the new layout packs each page by graph locality, so one read returns several edgelists that a traversal is about to need, at any vector size.

Exploiting this decoupled layout, \thiswork redesigns the vector update system.
First, we employ \emph{\vecpruning} to determine when to stop reading the full vectors during position seeking.
For each query, it selectively loads full vectors to identify only a sufficient set of close neighbors and fills the remaining neighbors with roughly found shortcuts.
This significantly reduces position-seeking overhead through lower I/O.

In addition, \layout enables efficient caching of edgelists in memory, avoiding the cache capacity pressure from the packed layout.
However, a compact edgelist cache is effective only when position-seeking traversals exhibit sufficient locality.
In GVS systems, such locality is largely shaped by the \navgraph, a sampled (e.g., 1\%) in-memory proximity graph that provides good entry points to reduce the traversal hops~\cite{pipeann, starling}.
Traversals from these entry points tend to reuse common graph paths, creating a compact hot set of edgelists that can be effectively cached.
Maintaining a fresh \navgraph is therefore key to preserving this locality.
If the \navgraph is omitted or remains unchanged over many insertions~\cite{freshdiskann,odinann,ipdiskann}, the entry points become increasingly distant from newly inserted vertices, dispersing traversal paths and reducing cache effectiveness.
\thiswork therefore keeps the \navgraph fresh through a lightweight update that reuses the traversal work already performed during position seeking.

With the above strategies reinforcing one another, \thiswork significantly reduces the overhead of position seeking, thereby boosting the performance of concurrent search and update operations on high-dimensional vector databases.
We benchmarked \thiswork on multiple large-scale high-dimensional datasets with state-of-the-art GVS update systems, and it improves insertion throughput by up to 3.97$\times$.
Even under such high insertion concurrency, it reduces search latency by up to 75.69\% while increasing throughput by up to 4.00$\times$.
Even on the low-dimensional dataset, \thiswork improves insertion throughput by up to 3.11$\times$ while maintaining the concurrent search performance.

%% file: sections/background.tex
\section{Background}
\label{sec:background}

\begin{figure}[t]
    \centering
    \includegraphics[width=.830\columnwidth]{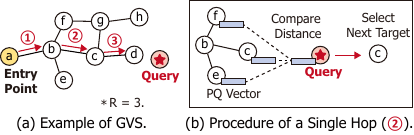}
    \Description{Diagram of a proximity graph search that starts at an entry point and greedily moves to closer neighbors until it reaches the vertex closest to the query.}
    \caption{Overview of graph-based vector search (GVS).}
    \label{fig:background1}
\end{figure}

\subsection{Graph-Based Vector Search Systems}
\label{sec:bg:gvs}

\begin{sloppypar}
Graph-based vector search (GVS) is the dominant index design for high-recall vector retrieval over large-scale vector datasets~\cite{nsg, hnsw, diskann, freshdiskann, odinann, pipeann}.
GVS organizes high-dimensional vectors into a \emph{proximity graph}~\cite{hnsw}, a sparse structure in which each vertex stores a vector and each edge links two vectors that are close under a distance metric such as L2.
A hyperparameter $R$ caps the per-vertex out-degree and trades exploration breadth for I/O and computation.
Existing frameworks typically set $R$ between 64 and 96 and fill each vertex up to $R$ to preserve graph quality.
\end{sloppypar}

\cref{fig:background1}(a) illustrates a search from an \emph{entry point} (vertex \texttt{a}) toward the closest vertex \texttt{d}.
At each hop, the query compares the distances of the current vertex's neighbors and greedily advances to the closest one (e.g., in \cref{fig:background1}(b), vertex \texttt{b} selects \texttt{c}), converging within a handful of hops.
Because exact distances over full-precision vectors are costly, GVS uses product-quantized (PQ) vectors~\cite{ivfadc} during traversal and defers exact distances to the visited vertices.

\subsection{On-Disk Graph-Based Vector Search}

\begin{figure}[t]
    \centering
    \includegraphics[width=.85\columnwidth]{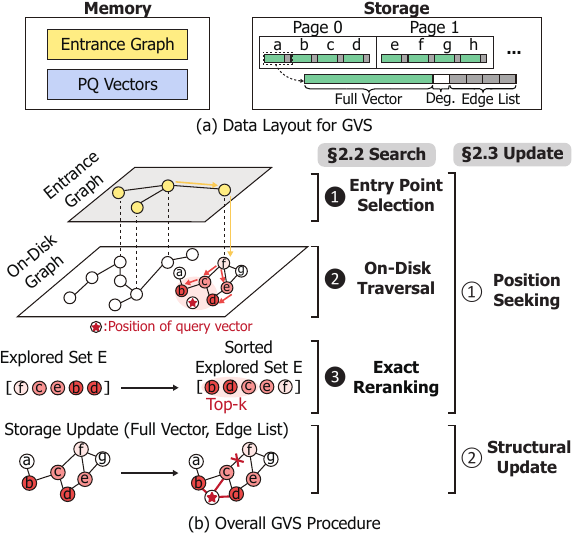}
    \Description{Diagram of the packed on-disk layout with entrance graph and PQ vectors in memory, and the search and insertion steps of position seeking and structural update.}
    \caption{On-disk GVS data layout, search, and update.}
    \label{fig:background2}
\end{figure}

In modern vector databases, the proximity graph from \cref{sec:bg:gvs} and its accompanying full vectors often far exceed host DRAM.
\emph{On-disk GVS systems}~\cite{diskann, starling, pipeann, qdrant} therefore offload both to SSD and load them on demand, providing a cost-effective path to high-recall search over millions of high-dimensional vectors~\cite{youtube,jdcom}.
As shown in \cref{fig:background2}(a), they keep a small in-memory \emph{\navgraph} and PQ vectors to make traversal cheap, while the full proximity graph and exact vectors are read selectively from the SSD.

\textbf{Packed storage layout.}
To minimize the number of storage I/Os, prior frameworks~\cite{diskann, starling, pipeann, qdrant} adopt a \emph{packed layout} in which a vector and its edge list reside contiguously within a 4KiB SSD page as a \texttt{[Vector] [Degree] [Edge List]} pair (\cref{fig:background2}(a)).
This packed layout lets the full vector needed by the final exact top-$K$ selection piggyback on the edgelist read, saving a second I/O.
For low-dimensional vectors, several pairs share one page, and this page-level locality further reduces the number of storage I/Os~\cite{starling}.

\textbf{Procedure of search.}
Using the packed layout, baseline on-disk GVS retrieves the top-$K$ nearest vectors in three stages~\cite{diskann, starling, pipeann}.

\bcircled{1} \emph{Entry-point selection}.
The system first searches the in-memory \navgraph, a sampled subset of the proximity graph with reduced connectivity (e.g., 1\% of vertices, $R$=32)~\cite{starling,pipeann}.
This process locates well-placed entry points for \bcircled{2}, reducing on-disk hops.

\bcircled{2} \emph{On-disk traversal}.
From these entry points, a greedy beam search over the on-disk graph advances toward closer neighbors using PQ distances, avoiding heavy computation and I/O overhead.
It keeps the current top-$|E_{search}|$ closest vertices in a fixed-size explored candidate pool (\texttt{Explored Set}) until no candidate yields a closer neighbor.
Although traversal never uses full vectors, each edgelist read under the packed layout brings the full vector along.

\bcircled{3} \emph{Exact reranking}.
Because PQ distance-based examinations are approximate, the system recomputes exact distances for the candidates in $E_{search}$ using their full vectors, which are already loaded into memory via piggybacking.
Using these exact distances, the system reranks $E_{search}$ and returns the final top-$K$ vectors.
However, many piggybacked vector loads are unnecessary during position seeking.
We will discuss this issue in \cref{sec:moti}.

\subsection{In-Place Update in On-Disk GVS Systems}
\label{sec:background3}

Production deployments also receive continuous updates~\cite{rag-update1, rag-update2, recom1, recom2, youtube}.
Among prevailing update-support strategies, we focus on in-place updates~\cite{odinann,ipdiskann}, which commit each insertion directly on disk to avoid the search-latency fluctuations of buffered merges~\cite{freshdiskann,greator} and quickly reflect fresh vectors into the index.
\cref{fig:background2}(b) shows $\star$ inserted into a graph of \texttt{a-g} in two steps, where \circled{1} position seeking finds the neighbors of $\star$ and \circled{2} structural update wires $\star$ to them.

\circled{1} \emph{Position seeking}.
This step finds neighbors that are semantically similar to the new vector to preserve graph quality, and it runs the same procedure as GVS search (\bcircled{1}--\bcircled{3}).
For instance, \texttt{b}, \texttt{c}, and \texttt{d} are closest to $\star$ within the explored set ($E_{pos}$) and become its neighbors.
However, position seeking needs a much larger explored set ($|E_{pos}| \gg |E_{search}|$), because it must gather enough candidates to fill the new vertex up to out-degree $R$ (usually 64--96) rather than only the top-$K$ (e.g., 10).
This leads to much heavier storage I/O than a typical search and interferes with concurrent searches.

\circled{2} \emph{Structural update}.
With the neighbors from \circled{1} in hand, the system connects the new vertex to them (e.g., $\star$ to \texttt{b}, \texttt{c}, \texttt{d}), prunes an existing edge (e.g., with \texttt{e}, dotted line) when a neighbor would exceed out-degree $R$, and flushes the modified pages to SSD.

%% file: sections/motivation.tex
\section{Motivational Analyses}
\label{sec:moti}

\begin{figure}[t]
    \centering
    \includegraphics[width=.9\columnwidth]{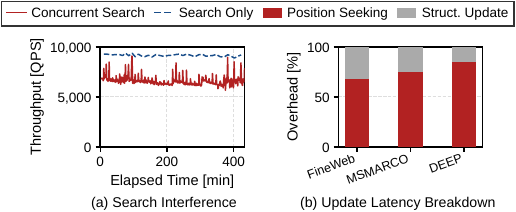}
    \Description{Bar charts comparing search-only and concurrent search throughput, and breaking down update latency into position seeking and structural update on three datasets.}
    \caption{Search interference and update-latency breakdown with OdinANN~\cite{odinann}.}
    \label{fig:moti_breakdown}
\end{figure}

\begin{sloppypar}

\subsection{Position Seeking: Index Update Bottleneck}

We first measure how index updates affect concurrent search.
\cref{fig:moti_breakdown}(a) compares search-only throughput with search throughput while the state-of-the-art framework (\ODANN~\cite{odinann}) inserts vectors into \msmarco under the default setup in \cref{sec:eval:setup}.
Search throughput drops by 27.89\% on average during the update.
\cref{fig:moti_breakdown}(b) separates update latency into position seeking and structural update on all three datasets.
Position seeking consumes up to 85\% of update time, making it the main bottleneck in search-update interference, and the following subsections identify its root causes.
\end{sloppypar}

\subsection{I/O Amplification from the Packed Layout}
\label{sec:moti:layout}

Position seeking constructs an $R$-neighbor list that contains close neighbors for local connectivity and long-range shortcuts for graph navigability~\cite{diskann,freshdiskann}.
The close neighbors must be distinguished accurately, while the shortcuts do not require the same distance accuracy.
Nevertheless, the packed layout reads every visited vertex's full vector because its edgelist cannot be read separately.

\begin{figure}[t]
    \centering
    \includegraphics[width=.87\columnwidth]{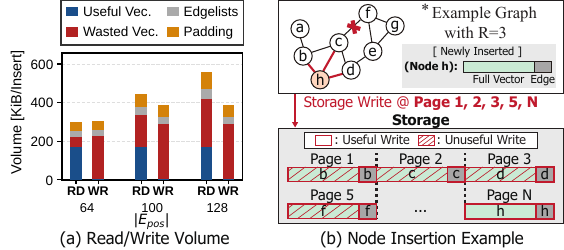}
    \Description{Stacked bars of bytes read and written per insertion split into useful vectors, wasted vectors, edgelists, and padding, with a diagram of pages rewritten during an insertion.}
    \caption{Read and write amplification from packed layout.}
    \label{fig:packed_limit_high_update}
\end{figure}

\cref{fig:packed_limit_high_update}(a) measures this extra I/O on \msmarco with $R=96$ while varying $|E_{pos}|$ from 64 to 128.
We divide the bytes read and written per insertion into \texttt{useful vector}, \texttt{wasted vector}, \texttt{edgelist}, and page-alignment \texttt{padding}.
Useful and wasted vectors are separated using \thiswork's PQ-distance-based classifier, which we describe in \cref{sec:pruning} and validate in \cref{sec:eval:casr}.
Once $|E_{pos}|$ exceeds $R$, the amount of \texttt{wasted vector} data increases with the explored set.
It reaches 44.34\% of all bytes read per insertion and consumes SSD bandwidth that could otherwise serve concurrent searches.
Additionally, packing loses its page-locality benefit for vectors larger than 2048 bytes because one vector and its edgelist leave no room for another record in a 4\,KiB page.

Moreover, during structural update, the packed layout unnecessarily rewrites full vectors.
As illustrated in \cref{fig:packed_limit_high_update}(b), inserting a vertex updates the edgelists of its $R$ neighbors, the pruned neighbor, and the new vertex itself, and each page update rewrites a full vector even though it is unchanged.
These rewrites account for up to 74.23\% of the bytes written per insertion (\cref{fig:packed_limit_high_update}(a)).

\subsection{Cache Pressure from the Packed Layout and Static \NavGraph}
\label{sec:moti:ent}

\textbf{Limited cache capacity.}
Position seeking explores far more vertices than top-$K$ search and creates a much larger edgelist working set, requiring careful cache management of reusable edgelists.
The packed layout significantly reduces the effective cache capacity because caching an edgelist also brings its co-located full vector into memory.
On high-dimensional datasets, these vectors consume most of the cached pages even though the cache aims to reuse edgelists.
Thus, the same memory budget holds fewer edgelists and covers a smaller fraction of the position-seeking working set.

\textbf{Locality loss from a static \navgraph.}
Before traversing the on-disk graph, the system searches the in-memory \navgraph to select its entry vertices.
A fresh \navgraph provides entry vertices that are close to the relevant regions of the current on-disk graph, which keeps the following traversal short.
Different traversals from these entry vertices often share their first few hops.
The edgelists on those common paths are accessed frequently and form the working set that an edgelist cache can retain.

\begin{figure}[t]
    \centering
    \includegraphics[width=.85\columnwidth]{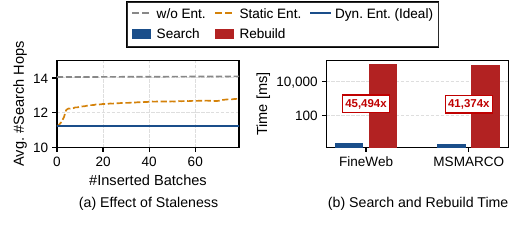}
    \Description{Line chart of average search hops as batches are inserted for no, static, and dynamic entrance graphs, and bar chart comparing search time with entrance graph rebuild time.}
    \caption{Staleness and rebuild cost of \navgraph.}
    \label{fig:in_mem_moti}
\end{figure}

Existing update-supported systems either omit the \navgraph or leave it unchanged during insertions to avoid its maintenance cost~\cite{odinann,freshdiskann}.
As new vertices are added, the fixed entry vertices become less effective at reaching the updated regions of the graph.
While inserting 0.8M vectors into a 0.5M-vector \imgnet index (\cref{fig:in_mem_moti}(a)), the static \navgraph requires more search hops as insertion proceeds while a dynamically refreshed one keeps traversal short.
Each additional hop accesses another edgelist and expands the working set that the cache must cover.
Rebuilding the entire \navgraph is not a practical solution, since one rebuild takes over 40,000$\times$ longer than a search (\cref{fig:in_mem_moti}(b))~\cite{diskann}.

\textbf{Decoupling or caching alone is insufficient.}
Separating vectors from edgelists lets the cache hold more edgelists, but removes the piggybacked vector read and can add SSD requests during exact reranking.
The separated layout therefore needs a policy that reads only the vectors required to select close neighbors.
A larger edgelist cache also provides little benefit when a static \navgraph spreads traversal over an expanding working set.
An effective solution must therefore coordinate the layout, vector-loading policy, \navgraph maintenance, and edgelist cache.

%% file: sections/overview.tex
\section{\thiswork Overview}
\label{sec:overview}

\begin{figure}[t]
    \centering
    \includegraphics[width=.981\columnwidth]{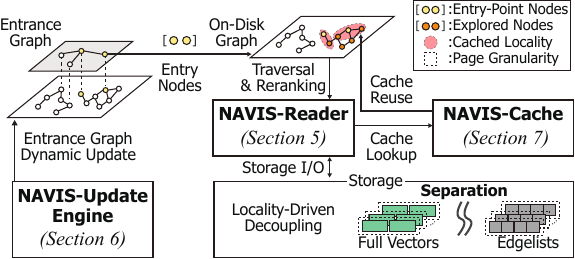}
    \Description{System diagram of NAVIS showing the decoupled on-disk storage, selective vector reranking, the dynamically updated entrance graph, and the edgelist cache.}
    \caption{Overview of \thiswork.
    }
    \label{fig:overview}
\end{figure}

As shown in \cref{sec:moti}, position seeking is now the main bottleneck during index update.
The packed layout amplifies read and write traffic and wastes cache capacity.
A static \navgraph lengthens on-disk traversal and expands the cache working set.
\thiswork (\cref{fig:overview}) addresses these problems with three components working together.

\textbf{\LayoutPruning (\cref{sec:layoutpruning}).}
\Layout stores full vectors and edgelists in separate on-disk files.
Traversal can thus read edgelists without fetching full vectors, while structural update writes only the changed records.
On this layout, \vecpruning selectively loads full vectors for close neighbors and fills the remaining edge slots with PQ-ranked shortcuts.

\textbf{\NavUpdate (\cref{sec:nav_update}).}
\thiswork keeps the \navgraph fresh by reusing the candidate sets that position seeking has already produced during insertions.
With that freshness, it keeps entry vertices close to newly inserted regions, shortening on-disk traversal and increasing the reuse of edgelists near those entry vertices.

\textbf{\CaChe (\cref{sec:cache}).}
The decoupled layout lets the cache devote memory to edgelists rather than co-located full vectors.
Our \navgraph keeps traversal paths short and concentrates accesses near the entry vertices.
\cache exploits the freed capacity and concentrated access locality by retaining edgelists from frequently accessed paths, while a small admission window filters out edgelists seen only during one-off exploration.

%% file: sections/method_layout.tex
\section{\LayoutPruning: Layout-Supported Selective Vector Reading}
\label{sec:layoutpruning}

\layoutpruning addresses the packed-layout limitations identified in \cref{sec:moti}.
It combines \layout, which separates full vectors from edgelists, with \vecpruning, which limits the additional vector requests introduced by separation.
Together, they reduce position-seeking I/O while retaining page-level locality.

\subsection{\LayOut}
\label{sec:layout}

\begin{figure}[t]
\centering
\includegraphics[width=.969\columnwidth]{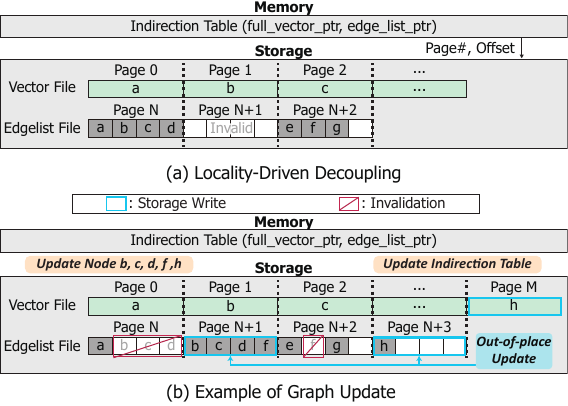}
\Description{Diagram of the decoupled layout with an edgelist file, a vector file, and an in-memory indirection table, and an example insertion that writes modified edgelists out of place.}
\caption{Overview of \layout.
}
\label{fig:unpack_overview}
\end{figure}

\begin{sloppypar}
To address the packed-layout inefficiencies, we propose \emph{\layout}, which separates full vectors from edgelists while reviving two optimizations originally designed for low-dimensional vectors.
For reads, it packs multiple edgelists into each 4\,KiB page to exploit page-level locality~\cite{starling}.
For writes, it updates modified edgelists out of place~\cite{odinann} without rewriting unchanged vectors.
\cref{fig:unpack_overview}(a) shows its three data structures.
The on-disk \textit{vector file} contains full-precision vectors, while the on-disk \textit{edgelist file} packs multiple edgelists into each page.
The memory-resident \emph{indirection table} locates these records by vertex ID.
\end{sloppypar}

The edgelist file in \layout enables page-level locality without being affected by the vector dimension because the decoupling lets edgelists gather densely without co-located vectors.
In \cref{fig:unpack_overview}(a), multiple edgelists share a single page regardless of vector size.
The nearby vertices' (\texttt{a}, \texttt{b}, \texttt{c}, and \texttt{d}) edgelists reside on one page, so reading \texttt{a}'s edgelist during traversal fetches \texttt{b}, \texttt{c}, and \texttt{d} in the same I/O, reducing the total traversal I/O count.

\Layout also enables updates with a much smaller write volume and fewer page writes.
As the vectors are decoupled from the edgelists, an update writes only the target vector to the vector file and skips the neighbors' vectors entirely.
In \cref{fig:unpack_overview}(b), only \texttt{h}'s own vector is written to the vector file, since unchanged neighbors' vectors are never touched.
\Layout also writes only the modified edgelists out of place, so inserting \texttt{h} gathers the edgelists of \texttt{b}, \texttt{c}, \texttt{d}, and \texttt{f} into a single new page.
Their old edgelist entries are invalidated, and the indirection table is updated to point to the new locations%
\footnote{The out-of-place update prefers writing to fully invalidated pages if possible, reducing overprovisioned storage. The new page for \texttt{b,c,d,f} reuses an invalidated page.}.
This reduces five full-page writes to two edgelist-page writes and one vector-file write, decoupling vector write volume from the out-degree $R$.

One concern is that page-level locality could be diminished during insertion.
Fortunately, the out-of-place write also preserves page-level locality through insertions.
Co-updated vertices tend to be adjacent in the graph and are likely to be co-traversed in future queries, so gathering them on a single new page maintains locality without explicit reorganization.
For the initial page-locality-aware placement, we utilize the same greedy placement as~\cite{starling}.

The indirection table costs host memory that grows only with the vertex count.
It keeps (vertex ID $\xrightarrow[]{}$ edge location) and (edge location $\xrightarrow[]{}$ vertex ID) mappings per vertex (two mappings $\times$ \#vertices $\times$ 4 Bytes), while vector locations are computed from the vertex ID rather than stored, which comes to 800MB at 100M vectors and 8GB at 1B.
The term is independent of vector dimension and out-degree, so it does not grow on the high-dimensional workloads that make position seeking expensive, and it matches what in-place baselines~\cite{odinann} already pay for out-of-place writes.

\subsection{\VecPruning}
\label{sec:pruning}

\input{algorithms/1_casr}

Realizing vector I/O savings on the decoupled layout requires \emph{\vecpruning}, which selectively loads vectors based on two key insights.
First, position seeking has a different goal than search.
Search must rerank a small explored set ($|E_{search}|$, typically 10--80) tightly to surface the exact top-$K$, so every candidate's exact distance matters.
Position seeking instead needs a much larger pool ($|E_{pos}| \gg |E_{search}|$) for thorough exploration, but only to surface adequate top-$R$ neighbors for the new vertex, following Vamana's~\cite{diskann} philosophy of combining close neighbors with long-range shortcuts.
Exhaustive reranking is therefore unnecessary.
Second, PQ distances computed during traversal are a strong signal for close neighbors, since candidates with the smallest PQ distances are most likely the closest under exact distances, which gives a reliable order for issuing full-vector loads.

\Vecpruning replaces the final reranking of position seeking, which conventionally loads full vectors for every candidate in $E_{pos}$.
Under typical settings ($|E_{pos}|=100$, $R=96$), nearly the entire pool becomes edges of the new vertex, yet only close neighbors require exact ranking.
We therefore load full vectors in PQ-distance order and track the exact top-$K$ among them, where $K$ is the search's top-$K$ size (e.g., $K=10$).
Once the top-$K$ stops changing, the close-neighbor portion is settled and the remaining $R$ edges fall to shortcut slots, so the number of exact-distance computations adapts to when each query's top-$K$ stabilizes.

\cref{alg:casr} implements this procedure with low I/O submission overhead, since issuing one I/O at a time per neighbor would be too costly.
It loads vectors in groups of size $s$ and pipelines each group's I/O submission with the exact-distance computation of the previous group, hiding submission latency through speculative loads (lines 6--9).
After each group, it updates the top-$K$ and compares it with the previous one (lines 10--16), terminating once they are identical and returning the exact-distance set $D$ and the converged top-$K$.

For position seeking, $D$ provides exact distances for the close-neighbor portion, and the new vertex's remaining $R$ edges are filled with other neighbors based on PQ distances (\cref{sec:navis_reader_example}), since shortcut slots only require diversity rather than exact closeness.
\Vecpruning also runs for search by substituting $E^{sorted}_{pos}$ with $E^{sorted}_{search}$ in \cref{alg:casr}, with the top-$K$ set $T$ used directly as the result.
The vector I/O reduction gain is smaller than for position seeking since the smaller $|E_{search}|$ leaves less room for early stopping, consistent with the sub-saturation regime ($|E_{pos}|=64$) in \cref{fig:packed_limit_high_update}(a).
We use different $s$ for the search and position-seeking paths because $|E_{pos}| \gg |E_{search}|$.

Group size $s$ is calibrated once at warm-up using 100 queries.
To obtain $s$, we initially set $s{=}1$ to get the distribution of the number of vectors needed for the top-$K$ to stabilize across all queries.
We then set $s$ to the P25 of the distribution, balancing convergence-detection granularity against I/O submission overhead.
We find P25 to be a good sweet spot, as extreme values are clearly suboptimal since $s{=}|E_{pos}|$ degenerates to a full fetch of $|E_{pos}|$ vectors and very small $s$ submits I/O too frequently, as shown in \cref{sec:eval:s_sensi}.

\begin{figure}[t]
\centering
\includegraphics[width=.9\columnwidth]{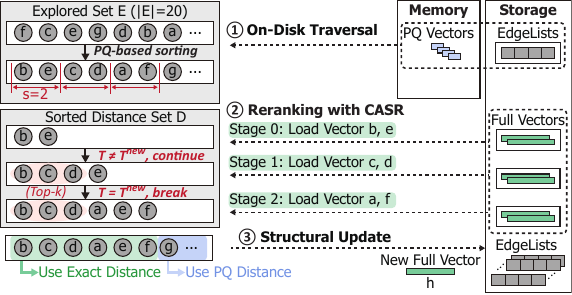}
\Description{Step-by-step example of an insertion with on-disk traversal, staged full-vector loading that stops when the top-K set stabilizes, and the structural update.}
\caption{Detailed example of \layoutpruning.}
\label{fig:detailed_example}
\end{figure}

\subsection{Update Example}
\label{sec:navis_reader_example}

\cref{fig:detailed_example} illustrates an update with \layoutpruning on the example graph of earlier figures, omitting entry-point selection for brevity.
\circled{1} \textbf{On-disk traversal.} Inserting a new vector traverses the on-disk graph and adds the visited vertices to $E_{pos}$ ($|E_{pos}|=20$ here, much larger in practice), reading only edgelist pages and comparing distances with host-memory PQ vectors, so \layout avoids full-vector I/O.
\circled{2} \textbf{Exact reranking with \vecpruning.}
Sorting $E_{pos}$ by PQ distance gives $E^{sorted}_{pos}=\{\texttt{b,e,c,d,a,f,g,...}\}$.
With $s=2$ and $K=3$, full vectors are loaded group by group, overlapping each group's I/O with the previous group's exact-distance computation, while the running top-$K$ is maintained.
The top-$K$ set $\{\texttt{b,c,d}\}$ stabilizes after Stage 2, so \vecpruning stops after 6 full-vector loads instead of 20, and the remaining $R-6$ neighbors fall to shortcut slots filled using deflated PQ vectors.
\circled{3} \textbf{Structural update.}
With the neighbor list assembled, the system writes only the new vertex's full vector and the modified edgelist pages, benefiting from \layout's decoupled writes.

%% file: algorithms/1_casr.tex
\begin{algorithm}[t]
\caption{\VecPruning (CASR).}
\label{alg:casr}
\small
\renewcommand{\algorithmicrequire}{\textbf{Input:}}
\renewcommand{\algorithmicensure}{\textbf{Output:}}
\begin{algorithmic}[1]
\Require Query $q$, candidate set sorted by PQ distance $E^{sorted}_{pos}$,
         top-$K$ size $K$, vector loading group size $s$ (calibrated at warm-up).
\Ensure Dict with $(\texttt{vertex\_id, exact distance})$ pairs ($D$), Top-$K$ neighbors of $q$ ($T$).

\State $D,\, T \gets \emptyset,\, \emptyset$ \Comment{Init exact distance set and top-$K$ set}
\State $idx \gets 0$ \Comment{Offset in $E^{sorted}_{pos}$ for vector loading}
\State $\text{VecLoad}(E^{sorted}_{pos},\, idx,\, s)$;\;  $idx += s$  \Comment{Load pipeline start}

\While{$\text{PrevVecLoad is non-empty}$}
    \State $\text{WaitPrevVecLoad}()$
    \If{$idx < |E^{sorted}_{pos}|$}                                          \Comment{Speculative next I/O}
        \State $s \gets \min(s,\, |E^{sorted}_{pos}| - idx)$
        \State $\text{VecLoad}(E^{sorted}_{pos},\, idx,\, s)$;\;  $idx += s$
    \EndIf
    \ForAll{$(\texttt{vertex\_id}, \texttt{vector}) \in \text{PrevVecLoad}$}
        \State $d \gets \text{L2Distance}(q,\, \texttt{vector})$
        \State $D \gets D \cup \{(\texttt{vertex\_id},\, d)\}$ \Comment{Exact distance compute/store}
    \EndFor
    \State $T^{new} \gets \text{TopK}(D,\, K)$ \Comment{Get new top-$K$ set from the updated $D$}
    \State{\textbf{if} $\text{Equal}(T,\, T^{new})$ \textbf{then} \textbf{break}}            \Comment{Convergence test}
    \State \textbf{else} $T \gets T^{new}$ \textbf{endif}
\EndWhile
\State \Return $(D, T)$
\end{algorithmic}
\end{algorithm}

%% file: sections/method_entrance.tex
\input{algorithms/2_nav_update}

\section{\NavUpdate: Dynamically Updating the Entrance Graph}
\label{sec:nav_update}

\Navupdate keeps the \navgraph fresh as the on-disk graph grows.
As insertions accumulate, the entry vertices from a static \navgraph become less effective, lengthening position seeking and expanding the edgelist working set.
Keeping the \navgraph fresh therefore shortens traversal and reduces the working set that \cache must cover.
Rebuilding the entire \navgraph is prohibitively expensive, as shown in \cref{sec:moti:ent}.

Incrementally adding a vector to the \navgraph presents its own challenge because finding its neighbors normally requires another graph traversal.
\Navupdate avoids this traversal by reusing work that the same insertion has already performed.
Entry-point selection produces the \navgraph explored set $E_{\mathrm{ent}}$, while on-disk position seeking produces the explored set $E_{pos}$.
After the on-disk insertion completes, \navupdate uses both sets to construct the new vector's \navgraph neighbor list.

\cref{alg:nav_update} triggers an update when the \navgraph size $|G_{ent}|$ falls below a threshold ratio of the on-disk graph size ($r_{ent} \times |G|$), with $r_{ent}=0.01$ following prior work~\cite{starling,pipeann} (line 1).
It first intersects $E_{pos}$ with $G_{\mathrm{ent}}$ to obtain $E_{\mathrm{inter}}$, which retains only vertices already in the \navgraph so that no vertex other than the new vector $q$ is added to $G_{\mathrm{ent}}$ (line 2).
It then builds $q$'s neighbor list from $E_{\mathrm{inter}}$ and $E_{\mathrm{ent}}$, capped at $R_{ent}$ entries and prioritizing $E_{\mathrm{inter}}$ because its vertices are typically closer to the query (lines 3--4).
Finally, $q$'s neighbors add reciprocal links to $q$, optionally pruning their lists to the maximum degree $R_{ent}$ (line 5).

Despite performing updates on the in-memory \navgraph, \navupdate incurs minimal lock overhead during insertion (lines 6--9).
In practice, adding a single vertex with a handful of neighbors into $G_{\mathrm{ent}}$ takes only microseconds, and only a small fraction of queries are promoted into the \navgraph.
Consequently, lock contention on $G_{ent}$ is negligible even with tens of concurrent insertion threads, and \navupdate introduces less than 1\% throughput overhead, as shown in \cref{sec:eval:insert_breakdown}.

%% file: algorithms/2_nav_update.tex
\begin{algorithm}[t]
\caption{\NavUpdate}
\small
\label{alg:nav_update}
\begin{algorithmic}[1]
\Require Query vector $q$, \navgraph $G_{\mathrm{ent}}$, \navgraph explored set $E_{\mathrm{ent}}$, on-disk explored set during position seeking $E_{pos}$, subgraph ratio $r_{\mathrm{ent}}$

\If{$|G_{\mathrm{ent}}| / |G| < r_{\mathrm{ent}}$}
    \Comment{$G_{\mathrm{ent}}$ needs more coverage}
    \State $E_{\mathrm{inter}} \gets E_{pos} \cap G_{\mathrm{ent}}$
    \Comment{Select nbr candidates only from $G_{\mathrm{ent}}$}
    \State $\mathcal{N}_{\mathrm{ent}} \gets
    \begin{cases}
        E_{\mathrm{inter}} \cup E_{\mathrm{ent}}[1:R - |E_{\mathrm{inter}}|], & |E_{\mathrm{inter}}| < R,\\[0pt]
        E_{\mathrm{inter}}[1:R], & |E_{\mathrm{inter}}| \ge R
    \end{cases}$
    \State $q$.nbr $\gets \mathcal{N}_{\mathrm{ent}}$
    \State \textbf{for} $p \in q.nbr: p.\hat{nbr} \gets \textrm{prune}(p.nbr \cup q)$
    \State $\text{Lock}(G_{\mathrm{ent}})$
    \State $G_{\mathrm{ent}} \gets G_{\mathrm{ent}} \cup q$
    \State \textbf{for} $p \in q.nbr: p.nbr \gets p.\hat{nbr}$
    \State $\text{Unlock}(G_{\mathrm{ent}})$
\EndIf
\end{algorithmic}
\end{algorithm}

%% file: sections/method_cache.tex
\section{\CaChe: Prioritized In-Memory Caching of the Proximity Graph}
\label{sec:cache}

\begin{figure}[t]
    \centering
    \includegraphics[width=.920\columnwidth]{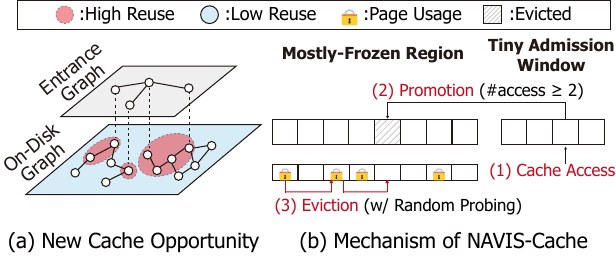}
    \Description{Diagram of edgelist access locality near entrance graph entry points and of the cache with a mostly-frozen region and a small admission window.}
    \caption{
    Overview of \cache.
    }
    \label{fig:cache_overview}
\end{figure}

The packed layout wastes cache capacity on co-located vectors.\footnote{Although an edgelist cache could conceptually operate at sub-page granularity, each update requires a page-wise read-modify-write to maintain storage consistency.
We therefore manage the cache at page granularity.}
\Layout removes these vector bytes, allowing the same memory budget to retain more edgelist pages.
Meanwhile, \navupdate keeps entry vertices fresh, shortening position-seeking paths and increasing edgelist reuse near those vertices.
\cref{fig:cache_overview}(a) illustrates the resulting caching opportunity from them.

A conventional LRU cache is still a poor fit for position seeking.
Its long traversals access many edgelist pages only once or a few times.
Admitting these low-reuse pages can evict frequently reused pages and lower the hit rate, as confirmed in \cref{sec:eval:cache_policy}.
Moreover, LRU updates recency state on every hit, adding bookkeeping and synchronization overhead under concurrent access~\cite{frozenhot}.

To exploit entry-point locality while avoiding LRU’s pitfalls, we design \emph{\cache}.
\cref{fig:cache_overview}(b) illustrates its structure.
Inspired by frozen-cache designs~\cite{frozenhot,tinylfu}, \cache is partitioned into a \emph{mostly-frozen region} (90\%) that preserves the hot region near entry points and protects it from eviction, and a \emph{tiny admission window} (10\%) that provides lightweight, usage-aware admission control.

An edgelist must be accessed at least twice within the admission window to be promoted into the mostly-frozen region.
The window itself is managed by LRU because of its substantially lower overhead compared to frequency-based policies such as LFU.
This rule effectively keeps hot edgelists near entry points while filtering out one-off accesses from long exploration paths, outperforming a full LRU cache as shown in the evaluation (\cref{sec:eval:cache_policy}).
As \navupdate refreshes the \navgraph during insertions, the admission window naturally incorporates newly relevant edgelists.

For the mostly-frozen region, \cache uses randomized eviction, selecting a random candidate and, if that entry is currently in use, performing a small number of random probes (up to eight by default) to find an alternative.
This avoids expensive tracking structures while preventing evictions of actively used entries.

%% file: sections/impl.tex
\section{Implementation}

\begin{sloppypar}
\textbf{I/O path optimization.}
\thiswork issues all storage I/O through \texttt{io\_uring}~\cite{io_uring}, which batches submissions and completions on per-thread rings and eliminates the per-request system-call overhead of synchronous read/write paths.
Because \naive use still incurs per-call kernel overhead under the large I/O bursts of position seeking, each worker thread owns a private ring of depth 256 and pre-registers both the edge and the vector files via \texttt{io\_uring\_register\_files}, so every submission carries an \texttt{IOSQE\_FIXED\_FILE} index instead of a raw file descriptor.
Each group of $s$ vector reads is committed in a single \texttt{io\_uring\_submit\_and\_wait} call and its completions are harvested in bulk via \texttt{io\_uring\_peek\_batch\_cqe}, paying a single enter-syscall per group and keeping this overhead near-constant.
\end{sloppypar}

\textbf{\Cache control details.}
Concurrent search and update force \cache to coexist with two write-side mechanisms, a read-modify-write (RMW) page cache that holds dirty pages until they are flushed, and \layout's out-of-place edge updates that relocate edgelists across pages.

\Cache is a write-through cache for clean read-side traffic, while only the RMW page cache holds dirty pages, so eviction from \cache never has to commit data and the RMW path never has to negotiate freshness with \cache.
On a search or position-seeking miss, the worker probes the RMW cache first to pick up any in-flight modifications, falls through to \cache for a clean hot page, and only issues storage I/O on a double miss, after which the page is admitted to the RMW cache and then to \cache's admission window.
Structural update operates exclusively on the RMW cache, pinning all dirty pages with reference counts and releasing the page lock only after the page is written to storage and any matching \cache entry is refreshed in place, an ordering that ensures \cache never holds a page newer than disk.

The out-of-place edge updates in \layout add a second consistency obligation, since an edgelist may move to a freshly allocated edge page, rendering the old slot stale.
We enforce freshness through the indirection table rather than through cache metadata.
A reader resolves \texttt{id}\,$\rightarrow$\,\texttt{edge\_list\_ptr} from the table before consulting any cache, so a hit on a relocated id targets the new page and the stale page is never queried again.
When all slots in a 4\,KiB edge page have been invalidated, the indirection layer issues an eviction hint to \cache for that page number, so the page is reclaimed before the cold-path randomized eviction would touch it.
Per-vertex tombstones and version counters are unnecessary, since invalidation is page-grained and the indirection table already provides the freshness guarantee.
This keeps \cache lock-free on the read path and free of dirty-page state management.

%% file: sections/eval.tex
\section{Evaluation}

\subsection{Experimental Setup}
\label{sec:eval:setup}

\textbf{Hardware.}
We used a single server for all evaluations, with an AMD Ryzen 9 7950X3D CPU (16C 32T, 4.2--5.7GHz), 128GB of DDR5 memory (4$\times$ DDR5 5600MT/s), one Micron Crucial PCIe 5.0 T705 4TB SSD, and Ubuntu 24.04 LTS as the operating system.

\begin{sloppypar}
\textbf{Datasets.}
We evaluated \thiswork on four datasets (\cref{tab:datasets}) and conducted the main experiments on three that span different dimensionalities and database scales.
\fineweb documents are embedded with EmbeddingGemma~\cite{embedding_gemma} and \imgnet images with CLIP~\cite{clip}, while \deep uses low-dimensional vectors generated by compression~\cite{deep} to show that \thiswork also performs well on low-dimensional datasets.
\imgnet is used only to evaluate a production baseline (\Qdrant~\cite{qdrant}) and a clustering-based baseline (\SPFresh~\cite{spfresh}) because neither reaches a comparable operating point on the high-dimensional, large-scale vector datasets of the main study within a practical time budget.
We inserted vectors equivalent to 20\%, 100\%, 20\%, and 30\% of the base index size for \fineweb, \msmarco, \deep, and \imgnet, respectively.
All datasets use $R = 96$ following prior work~\cite{diskann, odinann}, and the image-based vectors of \deep and \imgnet require much larger $|E_{pos}|$ values of 200 and 128 because they are highly similar to one another.
Following standard methodology~\cite{diskann, odinann}, we randomly selected 10{,}000 queries per dataset and targeted \texttt{Recall@10} of 0.95, 0.95, 0.85, and 0.95 for the four datasets, respectively, with $|E_{search}|$ = 40 for all, using a lower target for \deep because it has fewer dimensions than the others.

\end{sloppypar}

\input{tables/1_datasets}

\begin{sloppypar}
\textbf{Baselines.}
We compared \thiswork against four main GVS baselines and two additional baselines, for six baselines in total:
\begin{itemize}[leftmargin=*,itemsep=0pt,parsep=0pt,topsep=5pt]
\item \textbf{\FDANN}~\cite{freshdiskann}: a buffered-insertion design using a host-side buffer that is flushed and merged into the on-disk graph once the buffer reaches 6\% of the base graph size.
\item \textbf{\IPDANN}~\cite{ipdiskann}: a concurrent insertion baseline implemented in the current DiskANN3 library~\cite{diskann3}.
It performs in-place updates to prevent search fluctuation.
\item \textbf{\ODANN}~\cite{odinann}: a state-of-the-art in-place update that improves \IPDANN via better lock concurrency control, neighbor pruning during structural update, and overprovisioned writes.
\item \textbf{\ODANNP}: \ODANN with \cache. Since \ODANN retains the packed layout, \cache must cache packed pages (vectors co-resident with edgelists), so its effective edgelist cache size is smaller than \thiswork's for the same memory budget.
\item \textbf{\thiswork (Ours)}: a high-performance concurrent insertion framework that reduces unnecessary vector reads and employs a dynamic \navgraph with an efficient cache.
\end{itemize}

\end{sloppypar}

\begin{sloppypar}
All systems used a beamwidth of four.
Since \FDANN employs buffered insertion and merges the buffered graph whenever the index grows by 6\% of the base graph size, we derived its insertion throughput by averaging it over time.
We tested only \FDANN as the representative baseline for buffered-insertion approaches~\cite{freshdiskann,greator}, since these approaches are well known for concurrent search fluctuations under large updates~\cite{odinann}, which our evaluations also confirm.
Prior work~\cite{pipeann,starling} that does not support concurrent updates is excluded.
To test the limitations of current production and cluster-based systems, we also benchmarked \textbf{\Qdrant}~\cite{qdrant} and \textbf{\SPFresh}~\cite{spfresh}.
We report them only for \imgnet in \cref{sec:eval:s_sensi}, since neither reached a comparable operating point on the main datasets within a practical evaluation time.
\end{sloppypar}

\begin{figure*}[t]
    \centering
    \includegraphics[width=.980\textwidth]{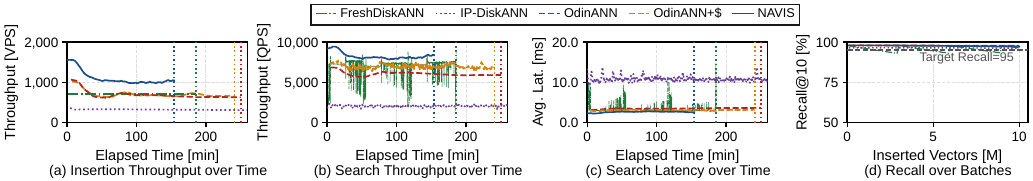}
    \Description{Time-series plots of insertion throughput, search throughput, search latency, and recall for all systems on the FineWeb dataset.}
    \caption{Update throughput and search performance under concurrent updates of baselines and \thiswork on the \fineweb dataset.}
    \label{fig:fineweb}
\end{figure*}

\begin{figure*}[t]
    \centering
    \includegraphics[width=.983\textwidth]{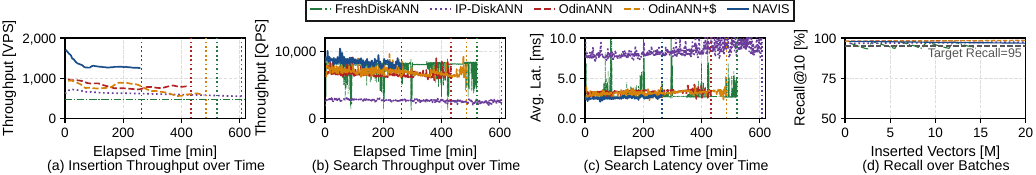}
    \Description{Time-series plots of insertion throughput, search throughput, search latency, and recall for all systems on the MSMARCO dataset.}
    \caption{Update throughput and search performance under concurrent updates of baselines and \thiswork on the \msmarco dataset.}
    \label{fig:msmarco}
\end{figure*}

\begin{figure*}[t]
    \centering
    \includegraphics[width=.979\textwidth]{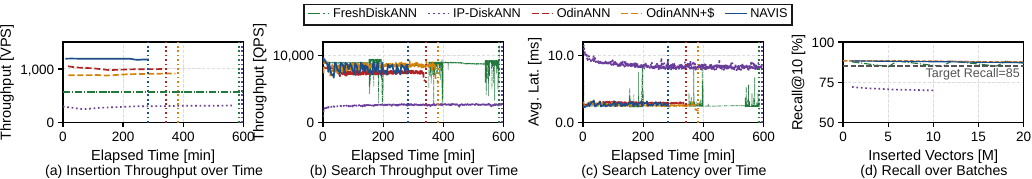}
    \Description{Time-series plots of insertion throughput, search throughput, search latency, and recall for all systems on the DEEP dataset.}
    \caption{Update throughput and search performance under concurrent updates of baselines and \thiswork on the \deep dataset.}
    \label{fig:deep}
\end{figure*}

\textbf{Other settings.}
We configured 22 search threads and 10 insertion threads by default, following prior work~\cite{odinann}.
Overall performance evaluations report up to 12 hours of insertions on slow baselines (i.e., \FDANN and \IPDANN), and other experiments report the first 100K insertions.
For search-only tasks, we warm up cache-based baselines by running the full 10{,}000-query search ten times.
\ODANNP and \thiswork use a 16GB host-memory cache for a fair comparison, matching the memory usage of \FDANN, whose insertion buffer takes much more host memory.
All GVS baselines enable the \navgraph with ten entry points.

\subsection{Concurrent Search-Insert Performance}
\label{sec:eval:overall}

In \cref{fig:fineweb}--\ref{fig:deep}, \thiswork delivers substantial improvements over the baselines, showing up to 3.97$\times$ higher insertion throughput.
Even under such high concurrent insertion, \thiswork achieves better search performance, up to 4.00$\times$ higher average queries per second (QPS), and up to 75.69$\%$ lower average latency, maintaining stable recall.
Its curves terminate earlier because the workload completes faster.

\begin{sloppypar}

\textbf{\fineweb.}
On \fineweb, \thiswork achieves 1.36/3.63/1.60/1.54$\times$ higher insertion throughput and 1.17/4.00/1.37/1.19$\times$ higher average QPS than \FDANN, \IPDANN, \ODANN, and \ODANNP, respectively, while reducing average latency by 18.98/75.69/25.26/15.06\%.
The \cache helps \ODANNP to some degree, but without the \layout, the performance improvement over \ODANN is limited, especially for insertion, due to redundant vector I/O and the small effective cache size from the packed layout.
\FDANN exhibits the buffered-merge instability previously reported by~\cite{odinann}, where its worst-case search QPS drops 79.1\% relative to its average during merge windows, visible as the spikes in \cref{fig:fineweb}(b).
\IPDANN avoids these fluctuations through in-place updates, but its concurrency control between searches and insertions limits both insertion and search throughput, which \ODANN addresses through improved lock concurrency control.
Recall fluctuates for \FDANN before each buffered batch is merged into the on-disk graph and recovers after the merge.
\thiswork, \ODANN, and \ODANNP converge to similar recall (near 97\%), all comfortably above the 95\% target.
\end{sloppypar}

\begin{sloppypar}

\textbf{\msmarco.}
On \msmarco, \thiswork improves insertion throughput by 2.74/2.15/1.63/1.83$\times$ and average QPS by 1.06/3.14/1.26/1.20$\times$, while reducing latency by 9.43/69.18/21.76/17.78\% over \FDANN, \IPDANN, \ODANN, and \ODANNP, respectively.
The search QPS gain over \FDANN is smaller than for \fineweb because \thiswork's insertion throughput on \msmarco far exceeds \FDANN's, interfering more with concurrent search.
Even so, \thiswork provides more stable and better search performance than \FDANN.
\ODANNP benefits from \cache for search, but due to the low hit rate from the small effective size with the packed layout, it shows lower insertion throughput than \ODANN.
Regarding recall, \FDANN again shows fluctuation (93\%--97\%) as in the \fineweb case, and \thiswork (97\%) matches \FDANN's max recall while sitting 1\% below \ODANN (98\%), still comfortably above the 95\% target.

\end{sloppypar}

\begin{sloppypar}

 \textbf{\deep (low-dimensional dataset test).}
Although \thiswork mainly targets high-dimensional vector databases, it remains competitive on \deep.
It improves insertion throughput by 2.07/3.96/1.21/1.35$\times$ over \FDANN, \IPDANN, \ODANN, and \ODANNP, respectively, while achieving comparable search throughput (0.93/3.11/1.08/0.96$\times$ QPS).
This advantage stems from \thiswork's reduced vector I/O and an optimized caching strategy that exploits the \navgraph locality.
\ODANNP confirms this, with higher search but lower insertion throughput than \ODANN since it does not reduce vector I/O.
\thiswork matches \ODANN's recall, while \FDANN shows the same merge-induced fluctuation.
\IPDANN reaches only 69.98\% under the same $|E_{search}|=40$ used by the other systems.

\end{sloppypar}

\subsection{Insert-Heavy Tests and Time Breakdown}
\label{sec:eval:insert_breakdown}

\begin{figure}[t]
    \centering
    \includegraphics[width=\columnwidth]{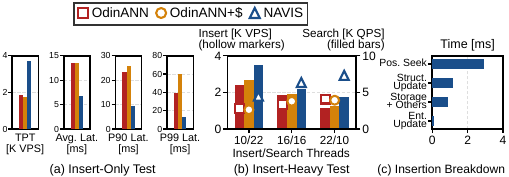}
    \Description{Bar charts of insert-only throughput and latency, insertion and search throughput across insert and search thread splits, and a breakdown of insertion time.}
    \caption{Insert-intensive tests and insert time breakdown.}
    \label{fig:insert_bench}
\end{figure}

We conducted insert-heavy benchmarks varying the \#insert/\#search thread ratio on \msmarco (\cref{fig:insert_bench}(a) and (b)), and broke down the insert time during the run in \cref{fig:insert_bench}(c).
In the insert-only test (i.e., 32/0 \#insert/\#search thread ratio), \thiswork's vector I/O reduction and effective cache yield 2.01/2.12$\times$ higher throughput than \ODANN and \ODANNP, respectively, while substantially reducing insert latency, especially on tail latencies (P90/P99, up to 65.74\% over \ODANN).
In the insert-heavy tests (\cref{fig:insert_bench}(b)) with the insert/search thread split over 10/22, 16/16, and 22/10, insertion throughput (hollow markers) over \ODANN rises from 1.64$\times$ at 10/22 to 1.79$\times$ at 22/10, since the position-seeking traffic that \thiswork reduces grows with the insert rate.
In \cref{fig:insert_bench}(c), \thiswork's vector I/O reduction shrinks the position-seeking share relative to \cref{fig:moti_breakdown} (74.83\% $\xrightarrow{}$ 58.24\%), and \texttt{Ent. Update}, the \navgraph update overhead of \navupdate, takes less than 1\% of the total time.

\subsection{Ablation Studies}
\label{sec:eval:ablation}
\label{sec:eval:casr}

\input{tables/2_casr_ablation}

\begin{figure}[t]
\centering
\includegraphics[width=.9\columnwidth]{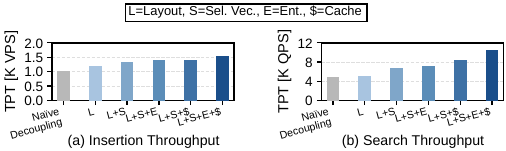}
\Description{Bar charts of insertion and search throughput as NAVIS components are added incrementally, starting from naive decoupling.}
\caption{Effect of \thiswork components.}
\label{fig:ablation}
\end{figure}

To evaluate the contribution of \thiswork's components, we conduct two complementary ablation studies.
\cref{tab:casr_results} first isolates \vecpruning (CASR) by comparing full reranking (\texttt{Full}) with CASR enabled (\texttt{+CASR}) while keeping all other \thiswork components unchanged.
To compare the recall accurately, we did not apply CASR for the concurrent search.
CASR reduces vector reads by up to 55.90\%, increasing insertion and search throughput by up to 1.85$\times$ and 1.59$\times$, respectively, while keeping the recall loss at most 0.58 pp.

\cref{fig:ablation} then builds \thiswork incrementally on \msmarco.
\textbf{\Naive Decoupling} uses our codebase and I/O path but only separates vectors and edgelists, without locality-aware placement, selective vector reads, \navgraph maintenance, or \cache.
\textbf{Layout (L)} adds locality-aware page placement, improving insertion/search throughput by 1.19/1.04$\times$, and \textbf{L+Sel. Vec. (L+S)} adds CASR for a further 1.10/1.32$\times$, showing that decoupling becomes substantially more effective with selective vector reads.
Adding \navupdate (\textbf{+Ent., L+S+E}) or \cache (\textbf{+\$, L+S+\$}) separately to L+S improves insertion/search throughput by 1.05/1.05$\times$ and 1.07/1.22$\times$, respectively, capturing only part of their joint benefit.
Adding \navupdate after \cache, as in \textbf{L+S+E+\$} (i.e., \thiswork), provides a further 1.08/1.26$\times$ over L+S+\$, raising the marginal search benefit of \navupdate from 1.05$\times$ without \cache to 1.26$\times$ with it.
This larger combined benefit shows that a fresh \navgraph and \cache reinforce each other, and the full system reaches an overall 1.51/2.11$\times$ insertion/search speedup over \Naive Decoupling.

Consistent with this interaction, \cref{fig:search_cache}(c) separately isolates \navupdate with \cache.
Removing \navupdate under the same cache policy lowers the cache hit rate by 6.78 pp, showing that a fresh \navgraph creates the stable locality that \cache exploits.
Applying the same cache to the packed layout yields only limited gains due to the smaller effective capacity for edgelists.
Together, these results show how \layout frees cache capacity, \navupdate creates stable locality, and \cache converts both into fewer SSD reads.

\subsection{Search Tail Latency Analysis}
\label{sec:eval:tail}

\begin{figure}[t]
\centering
\includegraphics[width=.924\columnwidth]{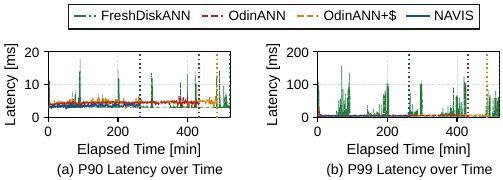}
\Description{Line plots of P90 and P99 search latency over time for the compared systems.}
\caption{Search tail latency (P90/P99) profiling results.}
\label{fig:tail_latency}
\end{figure}

\begin{sloppypar}
\cref{fig:tail_latency} shows the P90/P99 latency profile on \msmarco from the experiments in \cref{sec:eval:overall}.
\thiswork reduces the maximum P90 latency by 64.01\%/65.78\% relative to \FDANN and \ODANN, and the maximum P99 latency by 92.22\%/84.78\%, respectively.
This improvement primarily stems from \thiswork's reduction in vector I/O, which dominates long-latency outliers under concurrent updates.
\end{sloppypar}

\subsection{Memory and Storage Usage Analyses}
\label{sec:eval:mem_storage}

\begin{figure}[t]
    \centering
    \includegraphics[width=.95\columnwidth]{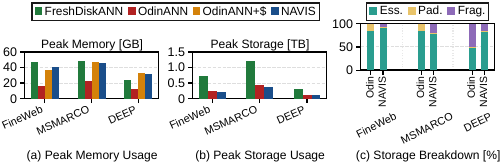}
    \Description{Bar charts of peak memory, peak storage, and a storage breakdown into essential bytes, padding, and fragmentation for each system and dataset.}
    \caption{Memory usage and storage usage with breakdown.}
    \label{fig:mem_storage}
\end{figure}

\begin{sloppypar}
\cref{fig:mem_storage} shows the peak memory consumption and storage usage of \thiswork compared to baselines.
\FDANN consumes more host memory than the others on high-dimensional datasets because of its host-memory merge buffer, so we configured \ODANNP and \thiswork with a 16GB \cache to match its peak, while \ODANN uses the least memory as it has no cache.
On \deep, \FDANN's host memory drops below \ODANNP and \thiswork because low-dimensional vectors yield a smaller insertion buffer, while both retain the 16GB \cache for consistency across datasets.
Regarding storage, \FDANN utilizes significantly more storage because it adopts a double-buffered storage scheme during merge.
\thiswork and \ODANN show much lower storage usage because they apply non-buffered updates, unlike \FDANN, with \thiswork slightly below \ODANN because it eliminates per-page padding.

\cref{fig:mem_storage}(c) breaks the storage footprint of both systems into essential bytes, padding, and fragmentation from out-of-place writes at the end of a full insert run.
\thiswork fragments only the edgelist file, and its allocator reuses fully invalidated pages when placing new writes, so on \deep its fragmentation stays at 15.52\% against \ODANN's 49.46\%, which fragments its packed pages.
What \ODANN avoids in fragmentation on \fineweb and \msmarco, it already pays in padding.
A page there holds a single vertex, so every rewrite invalidates the whole page and leaves no fragmentation behind, an outcome that follows from the padding those pages already carry.
The property does not carry over to \deep, where a page serves several vertices and fragmentation grows sharply.

\end{sloppypar}

\subsection{SSD Sensitivity Test, Search-Only Test, and Cache Policy Test}
\label{sec:eval:cache_policy}

\begin{figure}[t]
    \centering
    \includegraphics[width=.95\columnwidth]{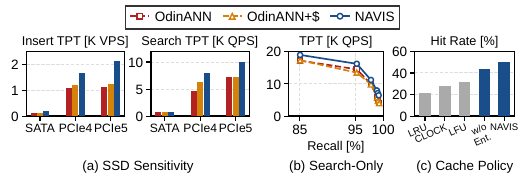}
    \Description{Bar charts of throughput on SATA, PCIe 4.0, and PCIe 5.0 SSDs, a search-only throughput versus recall curve, and cache hit rates under different cache policies.}
    \caption{SSD sensitivity, search-only test, and cache policies.}
    \label{fig:search_cache}
\end{figure}

To show that \thiswork consistently provides gains on more common SSDs, \cref{fig:search_cache}(a) repeats the concurrent search-insert test on \msmarco across a SATA SSD, a PCIe 4.0 NVMe SSD, and the PCIe 5.0 NVMe SSD used elsewhere.
On the more common PCIe 4.0 device, the gain over \ODANN persists at 1.71$\times$ for search and 1.53$\times$ for insertion.
SATA limits every system by the device, so we report it as an extreme end point rather than a deployment target.

While \thiswork primarily targets update overhead during position seeking, \layoutpruning and \cache also benefit search-only workloads, as shown in \cref{fig:search_cache}(b) on the \msmarco dataset with 32 search threads and zero insert threads.
Across all recall levels with various $|E_{search}|$, \thiswork provides higher QPS than \ODANN and \ODANNP, up to 1.60$\times$.

\cref{fig:search_cache}(c) tests \cache's policy choices on \fineweb with the cache forced to 4GB, since with \layout a 16GB cache yields over 90\% hit rate across all baselines.
We compare well-known LRU, FIFO with second chance (CLOCK), and LFU, and also \cache without \navupdate (\texttt{NAVIS-wo Ent}) to assess how \navupdate improves locality near entry points.
As expected, conventional policies suffer from cache pollution caused by low-reuse edgelists and fail to retain near-entry-point structures in the cache, whereas \cache effectively prevents such pollution and improves the hit rate by 19.28\,pp over LFU, and \navupdate adds another 6.78\,pp over the configuration without it.

\subsection{Group Size Sensitivity and Comparison with Other Baselines}
\label{sec:eval:s_sensi}

\begin{figure}[t]
    \centering
    \includegraphics[width=.92\columnwidth]{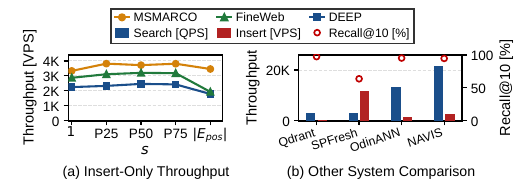}
    \Description{Line plots of insertion throughput versus group size on three datasets, and bar charts comparing Qdrant, SPFresh, OdinANN, and NAVIS on ImageNet.}
    \caption{Sensitivity of group size and comparison with production (\Qdrant) and cluster-based (\SPFresh) baselines.}
    \label{fig:other_baselines}
\end{figure}

Group size $s$ is calibrated per index rather than fixed across workloads, and the mechanism is insensitive to the exact value.
As described in \cref{sec:pruning}, $s$ is set at warm-up from the P25 of the per-query convergence distribution, with separate values for the search and position-seeking paths, so a new dataset needs no manual tuning.
In \cref{fig:other_baselines}(a), we tested the sensitivity of \thiswork to group size $s$ with the insert-only test using 32 threads on all three main datasets.
As expected, $s=1$ submits I/O too frequently and $s=|E_{pos}|$ loads all vectors in the candidate set, both showing lower insertion throughput than other $s$ choices.
The curves are flat across the middle of the range on every dataset, with the P25 to P75 spread within 5.71\%, so the P25 rule is a convenience rather than a requirement.
We also compared \thiswork against a production baseline (\Qdrant~\cite{qdrant}) and a clustering-based baseline (\SPFresh~\cite{spfresh}) on the \imgnet dataset in \cref{fig:other_baselines}(b).
For \Qdrant, we used the new HNSW~\cite{hnsw} features with storage from \texttt{version 1.16.3}, which use a strategy similar to \thiswork and the GVS baselines.
\Qdrant shows high recall, but its search and insert throughput are both significantly lower.
\SPFresh achieves high insert throughput from its batched, cluster-based design, but falls short on recall and search throughput.
Even on \imgnet, each baseline trails the GVS systems (i.e., \ODANN and \thiswork) in at least one of search throughput, insertion throughput, or recall, so we focus the main evaluation on GVS systems.

\subsection{Deletion Support}
\label{sec:eval:deletion}
\begin{figure}[t]
    \centering
    \includegraphics[width=.943\columnwidth]{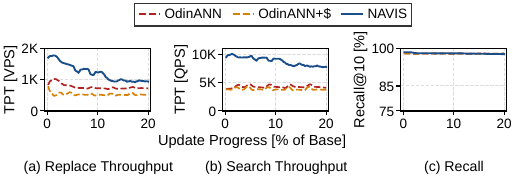}
    \Description{Plots of update throughput, search throughput, and recall over the course of a vector replacement benchmark.}
    \caption{Deletion benchmark by replacing the vectors.}
    \label{fig:deletion}
\end{figure}

\thiswork also supports deletion through the indirection table.
A deletion marks the vertex invalid in the table, searches skip invalid candidates using a slightly enlarged $E_{search}$, and a bulk consolidation runs once the invalidated fraction crosses a threshold, as in \ODANN.
Following the benchmark from~\cite{odinann, freshdiskann}, \cref{fig:deletion} runs a vector replacement (i.e., delete $\xrightarrow[]{}$ insert) benchmark on \fineweb, replacing 20\% of vectors.
\thiswork sustains 1.58/2.05$\times$ higher concurrent update/search throughput than \ODANN across the run, and \texttt{Recall@10} stays within 0.75\,pp of its starting value for all systems.
\ODANNP falls below \ODANN under this workload because deletions invalidate cached entries faster than the workload can replace them in a small effective cache size due to the packed layout, resulting in cache management overhead without benefit.

%% file: tables/1_datasets.tex
\begin{table}[t]
\centering
\caption{Dataset Configurations Used in NAVIS Experiments}
\label{tab:datasets}
\setlength{\tabcolsep}{5.3pt}
\resizebox{\columnwidth}{!}{%
\begin{tabular}{lccccccc}
\toprule
\textbf{Dataset} &
\makecell{\textbf{Dim.}\\\textbf{(float)}} &
\textbf{\#Vectors} &
\textbf{Type} &
\textbf{$|E_{search}|$} &
\textbf{$|E_{pos}|$} &
\makecell{\textbf{PQ}\\\textbf{Bytes}} \\
\midrule
\fineweb~\cite{fineweb_edu} & 768 & 60,000,000  & Text & 40 & 100 & 128 \\
\msmarco~\cite{msmarco} & 768 & 100,000,000 & Text & 40 & 100 & 128 \\
\deep~\cite{deep}    & 96  & 120,000,000 & Image & 40 & 200 & 32  \\
\imgnet~\cite{imagenet}  & 512 & 1,300,000   & Image & 40 & 128 & 64  \\
\bottomrule
\end{tabular}%
}
\end{table}

%% file: tables/2_casr_ablation.tex
\begin{table}[t]
\centering
\caption{Ablation Analysis of CASR}
\label{tab:casr_results}

\setlength{\tabcolsep}{11.4pt}
\resizebox{\columnwidth}{!}{%
\begin{tabular}{llrrrr}
\toprule
Dataset
& Type
& Insert TPT
& Search TPT
& \#Vec. Reads
& Recall@10 \\
\midrule

\multirow{2}{*}{\fineweb}
    & Full
    & 633.16 & 6016.33 & 103.31 & 97.73 \\
    & +CASR
    & 1171.57 & 9541.02 & 65.14 & 97.19 \\
\midrule

\multirow{2}{*}{\msmarco}
    & Full
    & 796.40 & 6705.56 & 104.30 & 98.43 \\
    & +CASR
    & 1408.89 & 8432.60 & 65.48 & 97.85 \\
\midrule

\multirow{2}{*}{\deep}
    & Full
    & 1002.37 & 7511.11 & 202.42 & 87.56 \\
    & +CASR
    & 1181.06 & 8096.12 & 89.27 & 87.45 \\

\bottomrule
\end{tabular}%
}
\end{table}

%% file: sections/discussion.tex
\section{Discussion}
\label{sec:discussion}

\textbf{Deployment conditions.}
The smaller gain on low-dimensional \deep shows that \thiswork's vector-I/O reduction is most beneficial on high-dimensional datasets, where avoided reads save more SSD traffic (\cref{tab:casr_results}).
The benefit grows with the update rate as more insertions place position-seeking traffic on the shared SSD (\cref{fig:insert_bench}(b)).
\thiswork remains advantageous across SSD classes, although the gain does not scale monotonically with device bandwidth because computation and synchronization also matter (\cref{fig:search_cache}(a)).
\cache is most advantageous under limited memory, because \layout enlarges effective cache capacity and our \navgraph stabilizes the hot set.
Its advantage over conventional policies narrows once the hot set fits in memory (\cref{fig:search_cache}(c)).

\textbf{Query-wise vector selection.}
Insertion queries vary in position-seeking difficulty, requiring different numbers of full-vector reads for their exact top-$K$ to stabilize.
\thiswork therefore uses query-wise adaptive vector selection.
To test this choice, we replace CASR with fixed-length vector selection used by another decoupled approach~\cite{dgai}, leaving the rest of \thiswork unchanged.
\cref{tab:vector_selection} reports this controlled comparison on \fineweb.
A fixed budget of 65 closely matches the average number of vector reads under CASR.
It achieves comparable insertion throughput and slightly higher concurrent search throughput by fetching all selected vectors in one I/O submission.
However, its recall is far below CASR and misses the 95\% target, showing that a fixed budget cannot handle the varying convergence behavior of insertion queries.

\input{tables/3_fixed_pruning}

\textbf{Application on emerging workloads.}
Agentic memory is an emerging use case for concurrent vector search and update.
Mem0 adds and consolidates memories during serving, while A-Mem links new memories and updates existing representations~\cite{mem0, amem}.
These updates must become searchable during retrieval, and they require stable serving/update throughput and latency.
When such stores grow beyond DRAM, \thiswork can provide an on-SSD index for this concurrent retrieval and update workload.

%% file: tables/3_fixed_pruning.tex
\begin{table}[t]
\centering
\caption{Fixed and Query-Adaptive Policy Comparison}
\label{tab:vector_selection}
\setlength{\tabcolsep}{16.8pt}
\resizebox{\columnwidth}{!}{%
\begin{tabular}{lrrrr}
\toprule
Method
& \#Vec. Reads
& Insert TPT
& Search TPT
& Recall@10 \\
\midrule

Fixed
& 30 & 1538.00 & 14301.61 & 92.53 \\

Fixed
& 65 & 1156.34 & 10868.60 & 93.64 \\

CASR
& 65.14 & 1171.57 & 9541.02 & 97.19 \\

\bottomrule
\end{tabular}%
}
\end{table}

%% file: sections/related_work.tex
\section{Related Work}
\label{sec:related}

\textbf{Graph-based vector databases.}
Graph-based vector search (GVS) spans many platforms~\cite{cagra, hnsw, starling, grasp, graphreorder_cache, finger, ganns, ggnn, pathweaver, song, pyramid, csdanns1, smartanns, df-gas, cxl-anns}.
On-disk systems~\cite{diskann, pipeann, hmann, filtered-diskann, ood-diskann, lm-diskann, csdanns1} provide low-cost scaling by storing the full graph on storage and compressed vectors in memory.
Recent systems~\cite{freshdiskann, odinann, dgai, ipdiskann, cleann} also support runtime graph updates.
FreshDiskANN~\cite{freshdiskann} buffers inserts and deletes, causing significant concurrent search performance fluctuations during merge.
IP-DiskANN~\cite{ipdiskann} introduced direct in-place updates; OdinANN~\cite{odinann} refines this model with approximate concurrency control and write overprovisioning.
We directly evaluate both in \cref{sec:eval:overall}.
CleanANN~\cite{cleann} reduces in-memory graph-update overhead, with in-memory rather than on-disk validation; its graph algorithm complements \thiswork.

\textbf{Decoupled on-disk layouts.}
Separating vectors from topology is not itself new.
DGAI~\cite{dgai} decouples vectors from the topology data and reduces redundant vector I/O during updates, but reranks a \textit{fixed} top-\(\tau\) candidate set selected at warm-up with enhanced PQ codebooks and keeps its \navgraph static.
\thiswork instead selects vectors per query until the exact top-\(K\) settles, refreshes the \navgraph, and exploits the resulting locality via \cache.
\cref{tab:vector_selection} and \cref{sec:discussion} (Query-wise vector selection) isolate this policy difference within \thiswork: fixed-length selection misses the target recall even with CASR-matched average vector reads.
DGAI's enhanced PQ compression remains complementary to these selective vector reads.
COMPASS~\cite{compass} (DecoupleVS in the latest preprint) decouples vector data and index metadata for component-specific lossless compression, with techniques to maintain comparable search and insertion throughput despite compression overhead.
It targets storage efficiency rather than position-seeking I/O, and its compression could be applied to \layout's vector file.

\begin{sloppypar}
\textbf{GPU-accelerated GVS.}
CAGRA~\cite{cagra} keeps the graph and FP32/FP16 vectors in GPU memory and co-designs graph construction and traversal for GPU parallelism.
BANG~\cite{bang} instead keeps the graph and full vectors in host memory, using GPU-resident PQ codes for traversal followed by exact reranking.
Both target static, memory-resident search rather than concurrent updates over an SSD-resident graph.
\thiswork supports large, high-dimensional datasets with an SSD-resident graph; GPUs can complement it by accelerating the remaining distance computations after it reduces position-seeking I/O.
\end{sloppypar}

\textbf{Concurrent search and update systems.}
Cluster-based methods~\cite{ivfadc, op_pq, re-ivf, scann, soar, spann, spfresh, faiss, reis, ice, quake} partition vectors, often with k-means, and localize updates instead of modifying multiple graph edges.
Recent dynamic systems use incremental rebalancing in SPFresh~\cite{spfresh}, temperature-based selective reindexing in Ada-IVF~\cite{ada-ivf}, and cost-based partition management in Quake~\cite{quake}.
Production systems~\cite{qdrant, milvus, manu, weaviate, vespa} support concurrent search and update.
\Qdrant~\cite{qdrant} pairs HNSW~\cite{hnsw} with on-disk storage; others avoid direct updates to SSD-resident graphs.
Milvus~\cite{milvus, manu} buffers vectors in growing in-memory segments, then seals them into immutable indexed segments, reproducing \FDANN's merge-window fluctuations.
Weaviate~\cite{weaviate} and Vespa~\cite{vespa} directly update mutable in-memory HNSW graphs, like in-memory libraries~\cite{faiss, hnsw}.
\cref{sec:eval:s_sensi} evaluates \Qdrant and \SPFresh.
On \imgnet, each trails the GVS systems in search throughput, insertion throughput, or recall, and neither reached a comparable operating point on our large high-dimensional datasets within a practical evaluation time.

%% file: sections/conclusion.tex
\section{Conclusion}

\begin{sloppypar}

This paper presents \thiswork, an on-disk GVS system enabling concurrent, high-performance search and update with low position-seeking overhead.
We identify position seeking as the main source of search-update interference, driven by uniform full-vector I/O and cache pressure in packed layouts and \navgraph staleness.
\thiswork mitigates these issues because decoupling enables selective vector reads and edgelist-only caching, while \navgraph maintenance preserves a stable hot set.
Across benchmarks, \thiswork achieves up to 3.97$\times$ higher insertion throughput, up to 4.00$\times$ higher search throughput, and up to 75.69\% lower search latency.

\end{sloppypar}

%% file: refs.bib
@misc{weaviate,
  author = {{Weaviate}},
  title = {{Weaviate}},
  year = {[n.d.]},
  url = {https://github.com/weaviate/weaviate},
  lastaccessed = {October 24, 2026}
}

@misc{vespa,
  author = {{Vespa}},
  title = {{Vespa}},
  year = {[n.d.]},
  url = {https://github.com/vespa-engine/vespa},
  lastaccessed = {October 24, 2026}
}

@ARTICLE{bang,
  author={Venkatasubba, Karthik and Khan, Saim and Singh, Somesh and Simhadri, Harsha Vardhan and Vedurada, Jyothi},
  journal={IEEE Transactions on Big Data (TBD)}, 
  title={{BANG: Billion-Scale Approximate Nearest Neighbor Search Using a Single GPU}}, 
  year={2025},
  volume={11},
  number={6},
  pages={3142-3157},
  doi={10.1109/TBDATA.2025.3581085}
  }

@inproceedings{mem0,
  author = {Chhikara, Prateek and Khant, Dev and Aryan, Saket and Singh, Taranjeet and Yadav, Deshraj},
  title = {{Mem0: Building Production-Ready AI Agents with Scalable Long-Term Memory}},
  booktitle = {Proceedings of the 28th European Conference on Artificial Intelligence (ECAI)},
  year = {2025},
  pages = {2993--3000},
  doi = {10.3233/FAIA251160}
}

@inproceedings{amem,
 author = {Xu, Wujiang and Liang, Zujie and Mei, Kai and Gao, Hang and Tan, Juntao and Zhang, Yongfeng},
booktitle = {Proceedings of the Advances in Neural Information Processing Systems 38 (NeurIPS)},
 doi = {10.52202/085713-0593},
 pages = {20004--20031},
 title = {{A-Mem: Agentic Memory for LLM Agents}},
 year = {2025}
}

@inproceedings{rag,
 author = {Lewis, Patrick and Perez, Ethan and Piktus, Aleksandra and Petroni, Fabio and Karpukhin, Vladimir and Goyal, Naman and K\"{u}ttler, Heinrich and Lewis, Mike and Yih, Wen-tau and Rockt\"{a}schel, Tim and Riedel, Sebastian and Kiela, Douwe},
 booktitle = {Proceedings of the Advances in Neural Information Processing Systems 33 (NeurIPS)},
 pages = {9459--9474},
 title = {{Retrieval-Augmented Generation for Knowledge-Intensive NLP Tasks}},
 url = {https://proceedings.neurips.cc/paper_files/paper/2020/file/6b493230205f780e1bc26945df7481e5-Paper.pdf},
 year = {2020}
}

@inproceedings{rec1,
author = {Huang, Zan and Chung, Wingyan and Ong, Thian-Huat and Chen, Hsinchun},
title = {{A graph-based recommender system for digital library}},
year = {2002},
doi = {10.1145/544220.544231},
booktitle = {Proceedings of the 2nd ACM/IEEE-CS Joint Conference on Digital Libraries (JCDL)},
pages = {65–73},
numpages = {9}
}

@ARTICLE{rec2,
  author={Li, Wen and Zhang, Ying and Sun, Yifang and Wang, Wei and Li, Mingjie and Zhang, Wenjie and Lin, Xuemin},
  journal={IEEE Transactions on Knowledge and Data Engineering (TKDE)}, 
  title={{Approximate Nearest Neighbor Search on High Dimensional Data — Experiments, Analyses, and Improvement}}, 
  year={2020},
  volume={32},
  number={8},
  pages={1475-1488},
  doi={10.1109/TKDE.2019.2909204}
}

@ARTICLE{multimedia1,
  author={Zhu, Lei and Zheng, Chaoqun and Guan, Weili and Li, Jingjing and Yang, Yang and Shen, Heng Tao},
  journal={IEEE Transactions on Knowledge and Data Engineering (TKDE)}, 
  title={{Multi-Modal Hashing for Efficient Multimedia Retrieval: A Survey}}, 
  year={2024},
  volume={36},
  number={1},
  pages={239-260},
  doi={10.1109/TKDE.2023.3282921}
}

@inproceedings{multimedia2,
author = {Huang, Jui-Ting and Sharma, Ashish and Sun, Shuying and Xia, Li and Zhang, David and Pronin, Philip and Padmanabhan, Janani and Ottaviano, Giuseppe and Yang, Linjun},
title = {{Embedding-based Retrieval in Facebook Search}},
year = {2020},
doi = {10.1145/3394486.3403305},
booktitle = {Proceedings of the 26th ACM SIGKDD International Conference on Knowledge Discovery \& Data Mining (KDD)},
pages = {2553–2561},
numpages = {9}
}

@article{embedding_gemma,
 author={Henrique Schechter Vera and Sahil Dua and Biao Zhang and Daniel Salz and Ryan Mullins and Sindhu Raghuram Panyam and Sara Smoot and Iftekhar Naim and Joe Zou and Feiyang Chen and Daniel Cer and Alice Lisak and Min Choi and Lucas Gonzalez and Omar Sanseviero and Glenn Cameron and Ian Ballantyne and Kat Black and Kaifeng Chen and Weiyi Wang and Zhe Li and Gus Martins and Jinhyuk Lee and Mark Sherwood and Juyeong Ji and Renjie Wu and Jingxiao Zheng and Jyotinder Singh and Abheesht Sharma and Divyashree Sreepathihalli and Aashi Jain and Adham Elarabawy and AJ Co and Andreas Doumanoglou and Babak Samari and Ben Hora and Brian Potetz and Dahun Kim and Enrique Alfonseca and Fedor Moiseev and Feng Han and Frank Palma Gomez and Gustavo Hernández Ábrego and Hesen Zhang and Hui Hui and Jay Han and Karan Gill and Ke Chen and Koert Chen and Madhuri Shanbhogue and Michael Boratko and Paul Suganthan and Sai Meher Karthik Duddu and Sandeep Mariserla and Setareh Ariafar and Shanfeng Zhang and Shijie Zhang and Simon Baumgartner and Sonam Goenka and Steve Qiu and Tanmaya Dabral and Trevor Walker and Vikram Rao and Waleed Khawaja and Wenlei Zhou and Xiaoqi Ren and Ye Xia and Yichang Chen and Yi-Ting Chen and Zhe Dong and Zhongli Ding and Francesco Visin and Gaël Liu and Jiageng Zhang and Kathleen Kenealy and Michelle Casbon and Ravin Kumar and Thomas Mesnard and Zach Gleicher and Cormac Brick and Olivier Lacombe and Adam Roberts and Qin Yin and Yunhsuan Sung and Raphael Hoffmann and Tris Warkentin and Armand Joulin and Tom Duerig and Mojtaba Seyedhosseini},
  title = {{EmbeddingGemma: Powerful and Lightweight Text Representations}},
  journal = {arXiv preprint arXiv:2509.20354},
  year = {2025},
  doi = {10.48550/arXiv.2509.20354}
}

@article{compass,
author={Yuanming Ren and Juncheng Zhang and Yanjing Ren and Rui Yang and Di Wu and Patrick P. C. Lee},
  title = {{Decoupling Vector Data and Index Storage for Space Efficiency}},
  journal = {arXiv preprint arXiv:2604.09173},
  year = {2026},
  doi = {10.48550/arXiv.2604.09173}
}

@InProceedings{clip,
  title = 	 {{Learning Transferable Visual Models From Natural Language Supervision}},
  author =       {Radford, Alec and Kim, Jong Wook and Hallacy, Chris and Ramesh, Aditya and Goh, Gabriel and Agarwal, Sandhini and Sastry, Girish and Askell, Amanda and Mishkin, Pamela and Clark, Jack and Krueger, Gretchen and Sutskever, Ilya},
  booktitle = 	 {Proceedings of the 38th International Conference on Machine Learning (ICML)},
  pages = 	 {8748--8763},
  year = 	 {2021},
  volume = 	 {139},
  url = 	 {https://proceedings.mlr.press/v139/radford21a.html}
}

@article{greator,
author = {Yu, Song and Lin, Shengyuan and Gong, Shufeng and Xie, Yongqing and Liu, Ruicheng and Zhou, Yijie and Sun, Ji and Zhang, Yanfeng and Li, Guoliang and Yu, Ge},
title = {{A Topology-Aware Localized Update Strategy for Graph-Based ANN Index}},
year = {2025},
volume = {19},
number = {3},
doi = {10.14778/3778092.3778108},
journal = {Proc. VLDB Endow. (PVLDB)},
pages = {495–508},
numpages = {14}
}

@inproceedings{recom1,
author = {Covington, Paul and Adams, Jay and Sargin, Emre},
title = {{Deep Neural Networks for YouTube Recommendations}},
year = {2016},
doi = {10.1145/2959100.2959190},
booktitle = {Proceedings of the 10th ACM Conference on Recommender Systems (RecSys)},
pages = {191–198},
numpages = {8}
}

@inproceedings{recom2,
author = {Eksombatchai, Chantat and Jindal, Pranav and Liu, Jerry Zitao and Liu, Yuchen and Sharma, Rahul and Sugnet, Charles and Ulrich, Mark and Leskovec, Jure},
title = {{Pixie: A System for Recommending 3+ Billion Items to 200+ Million Users in Real-Time}},
year = {2018},
doi = {10.1145/3178876.3186183},
booktitle = {Proceedings of the 2018 World Wide Web Conference (WWW)},
pages = {1775–1784},
numpages = {10}
}

@inproceedings{diskann,
 author = {Jayaram Subramanya, Suhas and Devvrit and Kadekodi, Rohan and Krishnaswamy, Ravishankar and Simhadri, Harsha Vardhan},
 booktitle = {Proceedings of Advances in Neural Information Processing Systems 32 (NeurIPS)},
 title = {{DiskANN: Fast Accurate Billion-point Nearest Neighbor Search on a Single Node}},
 url = {https://proceedings.neurips.cc/paper_files/paper/2019/file/09853c7fb1d3f8ee67a61b6bf4a7f8e6-Paper.pdf},
 volume = {32},
 year = {2019}
}

@article{nsg,
author = {Fu, Cong and Xiang, Chao and Wang, Changxu and Cai, Deng},
title = {{Fast approximate nearest neighbor search with the navigating spreading-out graph}},
year = {2019},
volume = {12},
number = {5},
doi = {10.14778/3303753.3303754},
journal = {Proc. VLDB Endow. (PVLDB)},
pages = {461–474},
numpages = {14}
}

@ARTICLE{hnsw,
  author={Malkov, Yu A. and Yashunin, D. A.},
  journal={IEEE Transactions on Pattern Analysis and Machine Intelligence (TPAMI)}, 
  title={{Efficient and Robust Approximate Nearest Neighbor Search Using Hierarchical Navigable Small World Graphs}}, 
  year={2020},
  volume={42},
  number={4},
  pages={824-836},
  doi={10.1109/TPAMI.2018.2889473}
}

@article{starling,
author = {Wang, Mengzhao and Xu, Weizhi and Yi, Xiaomeng and Wu, Songlin and Peng, Zhangyang and Ke, Xiangyu and Gao, Yunjun and Xu, Xiaoliang and Guo, Rentong and Xie, Charles},
title = {{Starling: An I/O-Efficient Disk-Resident Graph Index Framework for High-Dimensional Vector Similarity Search on Data Segment}},
year = {2024},
volume = {2},
number = {1},
doi = {10.1145/3639269},
journal = {Proc. ACM Manag. Data (PACMMOD)},
articleno = {14},
numpages = {27}
}

@inproceedings{milvus,
author = {Wang, Jianguo and Yi, Xiaomeng and Guo, Rentong and Jin, Hai and Xu, Peng and Li, Shengjun and Wang, Xiangyu and Guo, Xiangzhou and Li, Chengming and Xu, Xiaohai and Yu, Kun and Yuan, Yuxing and Zou, Yinghao and Long, Jiquan and Cai, Yudong and Li, Zhenxiang and Zhang, Zhifeng and Mo, Yihua and Gu, Jun and Jiang, Ruiyi and Wei, Yi and Xie, Charles},
title = {{Milvus: A Purpose-Built Vector Data Management System}},
year = {2021},
doi = {10.1145/3448016.3457550},
booktitle = {Proceedings of the 2021 International Conference on Management of Data (SIGMOD)},
pages = {2614–2627},
numpages = {14}
}

@inproceedings {pipeann,
author = {Hao Guo and Youyou Lu},
title = {{Achieving {Low-Latency} {Graph-Based} Vector Search via Aligning {Best-First} Search Algorithm with {SSD}}},
booktitle = {Proceedings of the 19th USENIX Symposium on Operating Systems Design and Implementation (OSDI)},
year = {2025},
pages = {171--186},
url = {https://www.usenix.org/conference/osdi25/presentation/guo}
}

@article{manu,
author = {Guo, Rentong and Luan, Xiaofan and Xiang, Long and Yan, Xiao and Yi, Xiaomeng and Luo, Jigao and Cheng, Qianya and Xu, Weizhi and Luo, Jiarui and Liu, Frank and Cao, Zhenshan and Qiao, Yanliang and Wang, Ting and Tang, Bo and Xie, Charles},
title = {{Manu: a cloud native vector database management system}},
year = {2022},
volume = {15},
number = {12},
doi = {10.14778/3554821.3554843},
journal = {Proc. VLDB Endow. (PVLDB)},
pages = {3548–3561},
numpages = {14}
}

@inproceedings {odinann,
author = {Hao Guo and Youyou Lu},
title = {{{OdinANN}: Direct Insert for Consistently Stable Performance in {Billion-Scale} {Graph-Based} Vector Search}},
booktitle = {Proceedings of the 24th USENIX Conference on File and Storage Technologies (FAST)},
year = {2026},
pages = {133--147},
url = {https://www.usenix.org/conference/fast26/presentation/guo},
}

@article{freshdiskann,
  author={Aditi Singh and Suhas Jayaram Subramanya and Ravishankar Krishnaswamy and Harsha Vardhan Simhadri},
  title = {{FreshDiskANN: A Fast and Accurate Graph-Based ANN Index for Streaming Similarity Search}},
  journal = {arXiv preprint arXiv:2105.09613},
  year = {2021},
  doi = {10.48550/arXiv.2105.09613}
}

@article{gorgeous,
author = {Yin, Peiqi and Yan, Xiao and Zhou, Qihui and Li, Hui and Li, Xiaolu and Wang, Meiling and Zhang, Lin and Yao, Xin and Cheng, James},
title = {{GPS: Revisiting the Data Layout for Disk-based High-Dimensional Vector Search}},
year = {2026},
volume = {4},
number = {3},
doi = {10.1145/3802069},
journal = {Proc. ACM Manag. Data (PACMMOD)},
articleno = {192},
numpages = {29}
}

@article{rag-update1,
  author={Jiale Han and Austin Cheung and Yubai Wei and Zheng Yu and Xusheng Wang and Bing Zhu and Yi Yang},
  title = {{RAG Meets Temporal Graphs: Time-Sensitive Modeling and Retrieval for Evolving Knowledge}},
  journal = {arXiv preprint arXiv:2510.13590},
  year = {2025},
  doi = {10.48550/arXiv.2510.13590}
}

@inproceedings{rag-update2,
    title = {{{D}ynamic{ER}: Resolving Emerging Mentions to Dynamic Entities for {RAG}}},
    author = {Kim, Jinyoung and Ko, Dayoon and Kim, Gunhee},
    booktitle = {Proceedings of the 2024 Conference on Empirical Methods in Natural Language Processing (EMNLP)},
    year = {2024},
    doi = {10.18653/v1/2024.emnlp-main.762},
    pages = {13752--13770},
}

@inproceedings{grasp,
author = {Zhang, Minjia and Wang, Wenhan and He, Yuxiong},
title = {{GraSP: Optimizing Graph-based Nearest Neighbor Search with Subgraph Sampling and Pruning}},
year = {2022},
doi = {10.1145/3488560.3498425},
booktitle = {Proceedings of the 15th ACM International Conference on Web Search and Data Mining (WSDM)},
pages = {1395–1405},
numpages = {11}
}

@inproceedings{graphreorder_cache,
 author = {Coleman, Benjamin and Segarra, Santiago and Smola, Alexander and Shrivastava, Anshumali},
 booktitle = {Proceedings of the Advances in Neural Information Processing Systems 35 (NeurIPS)},
 doi = {10.52202/068431-2789},
 pages = {38488--38500},
 title = {{Graph Reordering for Cache-Efficient Near Neighbor Search}},
 volume = {35},
 year = {2022}
}

@inproceedings{finger,
author = {Chen, Patrick and Chang, Wei-Cheng and Jiang, Jyun-Yu and Yu, Hsiang-Fu and Dhillon, Inderjit and Hsieh, Cho-Jui},
title = {{FINGER: Fast Inference for Graph-based Approximate Nearest Neighbor Search}},
year = {2023},
doi = {10.1145/3543507.3583318},
booktitle = {Proceedings of the ACM Web Conference 2023 (WWW)},
pages = {3225–3235},
numpages = {11}
}

@INPROCEEDINGS{song,
  author={Zhao, Weijie and Tan, Shulong and Li, Ping},
  booktitle={Proceedings of the IEEE 36th International Conference on Data Engineering (ICDE)}, 
  title={{SONG: Approximate Nearest Neighbor Search on GPU}}, 
  year={2020},
  pages={1033-1044},
  doi={10.1109/ICDE48307.2020.00094}
}

@INPROCEEDINGS{ganns,
  author={Yu, Yuanhang and Wen, Dong and Zhang, Ying and Qin, Lu and Zhang, Wenjie and Lin, Xuemin},
  booktitle={Proceedings of the IEEE 38th International Conference on Data Engineering (ICDE)}, 
  title={{GPU-accelerated Proximity Graph Approximate Nearest Neighbor Search and Construction}}, 
  year={2022},
  pages={552-564},
  doi={10.1109/ICDE53745.2022.00046}
}

@inproceedings{pyramid,
author = {Zhu, Zhenhua and Liu, Jun and Dai, Guohao and Zeng, Shulin and Li, Bing and Yang, Huazhong and Wang, Yu},
title = {{Processing-In-Hierarchical-Memory Architecture for Billion-Scale Approximate Nearest Neighbor Search}},
year = {2023},
doi = {10.1109/DAC56929.2023.10247946},
booktitle = {Proceedings of the 60th Annual ACM/IEEE Design Automation Conference (DAC)},
pages = {1–6},
numpages = {6}
}

@ARTICLE{csdanns1,
  author={Kim, Ji-Hoon and Park, Yeo-Reum and Do, Jaeyoung and Ji, Soo-Young and Kim, Joo-Young},
  journal={IEEE Transactions on Computers (TC)}, 
  title={{Accelerating Large-Scale Graph-Based Nearest Neighbor Search on a Computational Storage Platform}}, 
  year={2023},
  volume={72},
  number={1},
  pages={278-290},
  doi={10.1109/TC.2022.3155956}
}

@inproceedings {smartanns,
author = {Bing Tian and Haikun Liu and Zhuohui Duan and Xiaofei Liao and Hai Jin and Yu Zhang},
title = {{Scalable Billion-point Approximate Nearest Neighbor Search Using {SmartSSDs}}},
booktitle = {Proceedings of the 2024 USENIX Annual Technical Conference (USENIX ATC)},
year = {2024},
pages = {1135--1150},
url = {https://www.usenix.org/conference/atc24/presentation/tian}
}

@inproceedings{df-gas,
author = {Zeng, Shulin and Zhu, Zhenhua and Liu, Jun and Zhang, Haoyu and Dai, Guohao and Zhou, Zixuan and Li, Shuangchen and Ning, Xuefei and Xie, Yuan and Yang, Huazhong and Wang, Yu},
title = {{DF-GAS: a Distributed FPGA-as-a-Service Architecture towards Billion-Scale Graph-based Approximate Nearest Neighbor Search}},
year = {2023},
doi = {10.1145/3613424.3614292},
booktitle = {Proceedings of the 56th Annual IEEE/ACM International Symposium on Microarchitecture (MICRO)},
pages = {283–296},
numpages = {14}
}

@inproceedings {cxl-anns,
author = {Junhyeok Jang and Hanjin Choi and Hanyeoreum Bae and Seungjun Lee and Miryeong Kwon and Myoungsoo Jung},
title = {{{CXL-ANNS}: {Software-Hardware} Collaborative Memory Disaggregation and Computation for {Billion-Scale} Approximate Nearest Neighbor Search}},
booktitle = {Proceedings of the 2023 USENIX Annual Technical Conference (USENIX ATC)},
year = {2023},
pages = {585--600},
url = {https://www.usenix.org/conference/atc23/presentation/jang}
}

@ARTICLE{ggnn,
  author={Groh, Fabian and Ruppert, Lukas and Wieschollek, Patrick and Lensch, Hendrik P. A.},
  journal={IEEE Transactions on Big Data (TBD)}, 
  title={{GGNN: Graph-Based GPU Nearest Neighbor Search}}, 
  year={2023},
  volume={9},
  number={1},
  pages={267-279},
  doi={10.1109/TBDATA.2022.3161156}
}

@inproceedings {pathweaver,
author = {Sukjin Kim and Seongyeon Park and Si Ung Noh and Junguk Hong and Taehee Kwon and Hunseong Lim and Jinho Lee},
title = {{{PathWeaver}: A {High-Throughput} {Multi-GPU} System for {Graph-Based} Approximate Nearest Neighbor Search}},
booktitle = {Proceedings of the 2025 USENIX Annual Technical Conference (USENIX ATC)},
year = {2025},
pages = {1501--1517},
url = {https://www.usenix.org/conference/atc25/presentation/kim}
}

@INPROCEEDINGS{cagra,
  author={Ootomo, Hiroyuki and Naruse, Akira and Nolet, Corey and Wang, Ray and Feher, Tamas and Wang, Yong},
  booktitle={Proceedings of the IEEE 40th International Conference on Data Engineering (ICDE)}, 
  title={{CAGRA: Highly Parallel Graph Construction and Approximate Nearest Neighbor Search for GPUs}}, 
  year={2024},
  pages={4236-4247},
  doi={10.1109/ICDE60146.2024.00323}
}

@article{ipdiskann,
  author={Haike Xu and Magdalen Dobson Manohar and Philip A. Bernstein and Badrish Chandramouli and Richard Wen and Harsha Vardhan Simhadri},
  title = {{In-Place Updates of a Graph Index for Streaming Approximate Nearest Neighbor Search}},
  journal = {arXiv preprint arXiv:2502.13826},
  year = {2025},
  doi = {10.48550/arXiv.2502.13826}
}

@misc{diskann3,
  author = {{Microsoft}},
  title = {{DiskANN3: A Composable Vector Indexing Library}},
  year = {[n.d.]},
  url = {https://github.com/microsoft/DiskANN},
  lastaccessed = {October 24, 2026}
}

@inproceedings{cleann,
author = {Zhang, Ziyu and Wei, Yuanhao and Engels, Joshua and Shun, Julian},
title = {{CleanANN: Efficient and Robust Full Dynamism in Graph-based Approximate Nearest Neighbor Search}},
year = {2026},
doi = {10.1145/3816782.3819219},
booktitle = {Proceedings of the 38th ACM Symposium on Parallelism in Algorithms and Architectures (SPAA)},
pages = {247–260},
numpages = {14}
}

@inproceedings{hmann,
 author = {Ren, Jie and Zhang, Minjia and Li, Dong},
 booktitle = {Proceedings of the Advances in Neural Information Processing Systems 33 (NeurIPS)},
 pages = {10672--10684},
 title = {{HM-ANN: Efficient Billion-Point Nearest Neighbor Search on Heterogeneous Memory}},
 url = {https://proceedings.neurips.cc/paper_files/paper/2020/file/788d986905533aba051261497ecffcbb-Paper.pdf},
 volume = {33},
 year = {2020}
}

@inproceedings{filtered-diskann,
author = {Gollapudi, Siddharth and Karia, Neel and Sivashankar, Varun and Krishnaswamy, Ravishankar and Begwani, Nikit and Raz, Swapnil and Lin, Yiyong and Zhang, Yin and Mahapatro, Neelam and Srinivasan, Premkumar and Singh, Amit and Simhadri, Harsha Vardhan},
title = {{Filtered-DiskANN: Graph Algorithms for Approximate Nearest Neighbor Search with Filters}},
year = {2023},
doi = {10.1145/3543507.3583552},
booktitle = {Proceedings of the ACM Web Conference 2023 (WWW)},
pages = {3406–3416},
numpages = {11}
}

@article{ood-diskann,
  author={Shikhar Jaiswal and Ravishankar Krishnaswamy and Ankit Garg and Harsha Vardhan Simhadri and Sheshansh Agrawal},
  title = {{OOD-DiskANN: Efficient and Scalable Graph ANNS for Out-of-Distribution Queries}},
  journal = {arXiv preprint arXiv:2211.12850},
  year = {2022},
  doi = {10.48550/arXiv.2211.12850}
}

@INPROCEEDINGS{lm-diskann,
  author={Pan, Yu and Sun, Jianxin and Yu, Hongfeng},
  booktitle={Proceedings of the 2023 IEEE International Conference on Big Data (BigData)}, 
  title={{LM-DiskANN: Low Memory Footprint in Disk-Native Dynamic Graph-Based ANN Indexing}}, 
  year={2023},
  pages={5987-5996},
  doi={10.1109/BigData59044.2023.10386517}
}

@article{dgai,
  author={Jiahao Lou and Shufeng Gong and Quan Yu and Hao Guo and Youyou Lu and Song Yu and Yanfeng Zhang and Tiezheng Nie and Ge Yu},
  title = {{DGAI: Decoupled On-Disk Graph-Based ANN Index for Efficient Updates and Queries}},
  journal = {arXiv preprint arXiv:2510.25401},
  year = {2026},
  doi = {10.48550/arXiv.2510.25401}
}

@inproceedings{spann,
 author = {Chen, Qi and Zhao, Bing and Wang, Haidong and Li, Mingqin and Liu, Chuanjie and Li, Zengzhong and Yang, Mao and Wang, Jingdong},
 booktitle = {Proceedings of the Advances in Neural Information Processing Systems 34 (NeurIPS)},
 pages = {5199--5212},
 title = {{SPANN: Highly-efficient Billion-scale Approximate Nearest Neighborhood Search}},
 url = {https://proceedings.neurips.cc/paper_files/paper/2021/file/299dc35e747eb77177d9cea10a802da2-Paper.pdf},
 volume = {34},
 year = {2021}
}

@inproceedings{scann,
author = {Guo, Ruiqi and Sun, Philip and Lindgren, Erik and Geng, Quan and Simcha, David and Chern, Felix and Kumar, Sanjiv},
title = {{Accelerating large-scale inference with anisotropic vector quantization}},
year = {2020},
booktitle = {Proceedings of the 37th International Conference on Machine Learning (ICML)},
volume = {119},
pages = {3887--3896},
url = {https://proceedings.mlr.press/v119/guo20h.html}
}

@inproceedings{soar,
 author = {Sun, Philip and Simcha, David and Dopson, Dave and Guo, Ruiqi and Kumar, Sanjiv},
 booktitle = {Proceedings of the Advances in Neural Information Processing Systems 36 (NeurIPS)},
 doi = {10.52202/075280-0141},
 pages = {3189--3204},
 title = {{SOAR: Improved Indexing for Approximate Nearest Neighbor Search}},
 volume = {36},
 year = {2023}
}

@inproceedings{spfresh,
author = {Xu, Yuming and Liang, Hengyu and Li, Jin and Xu, Shuotao and Chen, Qi and Zhang, Qianxi and Li, Cheng and Yang, Ziyue and Yang, Fan and Yang, Yuqing and Cheng, Peng and Yang, Mao},
title = {{SPFresh: Incremental In-Place Update for Billion-Scale Vector Search}},
year = {2023},
doi = {10.1145/3600006.3613166},
booktitle = {Proceedings of the 29th Symposium on Operating Systems Principles (SOSP)},
pages = {545–561},
numpages = {17}
}

@article{ada-ivf,
  author={Jason Mohoney and Anil Pacaci and Shihabur Rahman Chowdhury and Umar Farooq Minhas and Jeffery Pound and Cedric Renggli and Nima Reyhani and Ihab F. Ilyas and Theodoros Rekatsinas and Shivaram Venkataraman},
  title = {{Incremental IVF Index Maintenance for Streaming Vector Search}},
  journal = {arXiv preprint arXiv:2411.00970},
  year = {2024},
  doi = {10.48550/arXiv.2411.00970}
}

@article{faiss,
  author = {Douze, Matthijs and Guzhva, Alexandr and Deng, Chengqi and Johnson, Jeff and Szilvasy, Gergely and Mazaré, Pierre-Emmanuel and Lomeli, Maria and Hosseini, Lucas and Jégou, Hervé},
  title = {{The Faiss Library}},
  journal = {IEEE Transactions on Big Data (TBD)},
  year = {2026},
  volume = {12},
  number = {2},
  pages = {346--361},
  doi = {10.1109/TBDATA.2025.3618474}
}

@ARTICLE{ivfadc,
  author={Jégou, Herve and Douze, Matthijs and Schmid, Cordelia},
  journal={IEEE Transactions on Pattern Analysis and Machine Intelligence (TPAMI)}, 
  title={{Product Quantization for Nearest Neighbor Search}}, 
  year={2011},
  volume={33},
  number={1},
  pages={117-128},
  doi={10.1109/TPAMI.2010.57}
}

@article{op_pq,
author = {Ge, Tiezheng and He, Kaiming and Ke, Qifa and Sun, Jian},
title = {{Optimized Product Quantization}},
year = {2014},
volume = {36},
number = {4},
doi = {10.1109/TPAMI.2013.240},
journal = {IEEE Transactions on Pattern Analysis and Machine Intelligence (TPAMI)},
pages = {744–755},
numpages = {12}
}

@inproceedings{re-ivf,
author = {Baranchuk, Dmitry and Babenko, Artem and Malkov, Yury},
title = {{Revisiting the Inverted Indices for Billion-Scale Approximate Nearest Neighbors}},
year = {2018},
doi = {10.1007/978-3-030-01258-8_13},
booktitle = {Proceedings of the 15th European Conference on Computer Vision (ECCV)},
pages = {209–224},
numpages = {16}
}

@inproceedings{reis,
author = {Chen, Kangqi and Nadig, Rakesh and Frouzakis, Manos and Ghiasi, Nika Mansouri and Liang, Yu and Mao, Haiyu and Park, Jisung and Sadrosadati, Mohammad and Mutlu, Onur},
title = {{REIS: A High-Performance and Energy-Efficient Retrieval System with In-Storage Processing}},
year = {2025},
doi = {10.1145/3695053.3731116},
booktitle = {Proceedings of the 52nd Annual International Symposium on Computer Architecture (ISCA)},
pages = {1171–1192},
numpages = {22}
}

@inproceedings{ice,
author = {Hu, Han-Wen and Wang, Wei-Chen and Chang, Yuan-Hao and Lee, Yung-Chun and Lin, Bo-Rong and Wang, Huai-Mu and Lin, Yen-Po and Huang, Yu-Ming and Lee, Chong-Ying and Su, Tzu-Hsiang and Hsieh, Chih-Chang and Hu, Chia-Ming and Lai, Yi-Ting and Chen, Chung-Kuang and Chen, Han-Sung and Li, Hsiang-Pang and Kuo, Tei-Wei and Chang, Meng-Fan and Wang, Keh-Chung and Hung, Chun-Hsiung and Lu, Chih-Yuan},
title = {{ICE: An Intelligent Cognition Engine with 3D NAND-Based In-Memory Computing for Vector Similarity Search Acceleration}},
year = {2022},
doi = {10.1109/MICRO56248.2022.00058},
booktitle = {Proceedings of the 55th Annual IEEE/ACM International Symposium on Microarchitecture (MICRO)},
pages = {763–783},
numpages = {21}
}

@inproceedings {quake,
author = {Jason Mohoney and Devesh Sarda and Mengze Tang and Shihabur Rahman Chowdhury and Anil Pacaci and Ihab F. Ilyas and Theodoros Rekatsinas and Shivaram Venkataraman},
title = {{Quake: Adaptive Indexing for Vector Search}},
booktitle = {Proceedings of the 19th USENIX Symposium on Operating Systems Design and Implementation (OSDI)},
year = {2025},
pages = {153--169},
url = {https://www.usenix.org/conference/osdi25/presentation/mohoney}
}

@misc{youtube,
  author = {{Google}},
  title = {{YouTube for Press}},
  year = {[n.d.]},
  url = {https://blog.youtube/press/},
  lastaccessed = {October 24, 2026}
}

@misc{io_uring,
  author = {Jens Axboe},
  title = {{Efficient IO with io\_uring}},
  year = {2019},
  url = {https://web.archive.org/web/20251229090652/https://kernel.dk/io_uring.pdf},
  lastaccessed = {October 24, 2026}
}

@misc{imagenet,
  author = {Stanford Vision Lab},
  title = {{ImageNet}},
  year = {[n.d.]},
  url = {https://www.image-net.org/index.php},
  lastaccessed = {October 24, 2026}
}

@misc{fineweb_edu,
    author       = {Lozhkov, Anton and Ben Allal, Loubna and von Werra, Leandro and Wolf, Thomas},
    title        = {{FineWeb-Edu: the Finest Collection of Educational Content}},
    year         = {2024},
    url          = {https://huggingface.co/datasets/HuggingFaceFW/fineweb-edu},
    doi          = {10.57967/hf/2497},
    publisher    = {Hugging Face},
    lastaccessed = {October 24, 2026}
}

@article{msmarco,
  author={Payal Bajaj and Daniel Campos and Nick Craswell and Li Deng and Jianfeng Gao and Xiaodong Liu and Rangan Majumder and Andrew McNamara and Bhaskar Mitra and Tri Nguyen and Mir Rosenberg and Xia Song and Alina Stoica and Saurabh Tiwary and Tong Wang},
  title = {{MS MARCO: A Human Generated MAchine Reading COmprehension Dataset}},
  journal = {arXiv preprint arXiv:1611.09268},
  year = {2018},
  doi = {10.48550/arXiv.1611.09268}
}

@inproceedings{jdcom,
author = {Li, Jie and Liu, Haifeng and Gui, Chuanghua and Chen, Jianyu and Ni, Zhenyuan and Wang, Ning and Chen, Yuan},
title = {{The Design and Implementation of a Real Time Visual Search System on JD E-commerce Platform}},
year = {2018},
doi = {10.1145/3284028.3284030},
booktitle = {Proceedings of the 19th International Middleware Conference Industry (Middleware)},
pages = {9–16},
numpages = {8}
}

@inproceedings{frozenhot,
author = {Qiu, Ziyue and Yang, Juncheng and Zhang, Juncheng and Li, Cheng and Ma, Xiaosong and Chen, Qi and Yang, Mao and Xu, Yinlong},
title = {{FrozenHot Cache: Rethinking Cache Management for Modern Hardware}},
year = {2023},
doi = {10.1145/3552326.3587446},
booktitle = {Proceedings of the 18th European Conference on Computer Systems (EuroSys)},
pages = {557–573},
numpages = {17}
}

@article{tinylfu,
author = {Einziger, Gil and Friedman, Roy and Manes, Ben},
title = {{TinyLFU: A Highly Efficient Cache Admission Policy}},
year = {2017},
volume = {13},
number = {4},
doi = {10.1145/3149371},
journal = {ACM Transactions on Storage (TOS)},
articleno = {35},
numpages = {31}
}

@misc{openai_embedding,
  author = {{OpenAI}},
  title = {{New embedding models and API updates}},
  year = {2024},
  url = {https://openai.com/index/new-embedding-models-and-api-updates/},
  lastaccessed = {October 24, 2026}
}

@misc{ada002,
  author = {{OpenAI}},
  title = {{Vector Embeddings -- OpenAI API}},
  year = {[n.d.]},
  url = {https://developers.openai.com/api/docs/guides/embeddings},
  lastaccessed = {October 24, 2026}
}

@inproceedings{bge_m3,
  author = {Chen, Jianlyu and Xiao, Shitao and Zhang, Peitian and Luo, Kun and Lian, Defu and Liu, Zheng},
  title = {{M3-Embedding: Multi-Linguality, Multi-Functionality, Multi-Granularity Text Embeddings Through Self-Knowledge Distillation}},
  booktitle = {Findings of the Association for Computational Linguistics (ACL Findings)},
  year = {2024},
  pages = {2318--2335},
  doi = {10.18653/v1/2024.findings-acl.137}
}

@misc{qdrant,
  author = {{Qdrant}},
  title = {{Qdrant: High-Performance Vector Search at Scale}},
  year = {[n.d.]},
  url = {https://qdrant.tech/},
  lastaccessed = {October 24, 2026}
}

@INPROCEEDINGS{deep,
  author={Babenko, Artem and Lempitsky, Victor},
  booktitle={Proceedings of the 2016 IEEE Conference on Computer Vision and Pattern Recognition (CVPR)}, 
  title={{Efficient Indexing of Billion-Scale Datasets of Deep Descriptors}}, 
  year={2016},
  pages={2055-2063},
  doi={10.1109/CVPR.2016.226}
}
